\input harvmac
\input psfig
\newcount\figno
\figno=0
\def\fig#1#2#3{
\par\begingroup\parindent=0pt\leftskip=1cm\rightskip=1cm\parindent=0pt
\global\advance\figno by 1
\midinsert
\epsfxsize=#3
\centerline{\epsfbox{#2}}
\vskip 12pt
{\bf Fig. \the\figno:} #1\par
\endinsert\endgroup\par
}
\def\figlabel#1{\xdef#1{\the\figno}}
\def\encadremath#1{\vbox{\hrule\hbox{\vrule\kern8pt\vbox{\kern8pt
\hbox{$\displaystyle #1$}\kern8pt}
\kern8pt\vrule}\hrule}}

\overfullrule=0pt

%
\def\tilde{\widetilde}
\def\bar{\overline}
\def\Z{{\bf Z}}

\def\S{{\bf S}}
\def\R{{\bf R}}

\font\zfont = cmss10 
\font\litfont = cmr6

\def\bigone{\hbox{1\kern -.23em {\rm l}}}
\def\ZZ{\hbox{\zfont Z\kern-.4emZ}}
\def\half{{\litfont {1 \over 2}}}

\Title{hep-th/9802150, IASSNS-HEP-98-15}
{\vbox{\centerline{ANTI DE SITTER SPACE}
\bigskip
\centerline{AND HOLOGRAPHY }}}
\smallskip
\centerline{Edward Witten}
\smallskip
\centerline{\it School of Natural Sciences, Institute for Advanced Study}
\centerline{\it Olden Lane, Princeton, NJ 08540, USA}\bigskip

\medskip

\def\N{{\cal N}}
\noindent
Recently, it has been proposed by Maldacena that large $N$ limits of certain
conformal field theories in $d$ dimensions
can be described in terms of supergravity (and string theory) on the product of
$d+1$-dimensional $AdS$ space with a compact manifold.  
Here we elaborate on this idea and propose a precise correspondence
between conformal field theory observables and those of supergravity:
correlation functions in conformal field theory are given by the
dependence of the supergravity action on the asymptotic behavior at
infinity.  In particular, dimensions of operators in conformal field theory
are given by masses of particles in supergravity.  As quantitative confirmation
of this correspondence, we note that the Kaluza-Klein modes of
Type IIB supergravity on $AdS_5\times {\bf S}^5$ match 
with the chiral operators of $\N=4$ super Yang-Mills theory in
four dimensions.  With some further assumptions, one can deduce a
Hamiltonian version of the correspondence and show that the
$\N=4$ theory has a large $N$ phase transition related to the thermodynamics
of $AdS$ black holes. 

\Date{February, 1998}

\newsec{Introduction}

To understand the large $N$ behavior of gauge theories with $SU(N)$
gauge group is a longstanding problem \ref\thooft{G. 't Hooft, ``A Planar 
Diagram Theory For Strong Interactions,'' Nucl. Phys. {\bf B72} (1974) 461.},
and offers perhaps the best hope of eventually understanding the
classic strong coupling mysteries of QCD.  
It has long been suspected that the large $N$ behavior, if accessible at all,
should be described by string theory perhaps with Liouville fields and
higher dimensions; see \ref\polyakov{A. M. Polyakov, ``String Theory
And Quark Confinement,'' hep-th/9711002.}  for recent discussion.
Lately, with hints coming from explorations of the near horizon structure
of certain black hole metrics 
\nref\gibo{G. Gibbons, Nucl. Phys. {\bf B207} (1982) 337; R. Kallosh
and A. Peet, Phys. Rev. {\bf B46} (1992) 5223, hep-th/9209116; S. Ferrara,
G. Gibbons, R. Kallosh, Nucl. Phys. {\bf B500} (1997) 75, hep-th/9702103.}
\nref\gibtot{G. Gibbons and P. Townsend, ``Vacuum Interpolation In Supergravity
Via Super $p$-Branes,'' Phys. Rev. Lett. {\bf 71} (1993) 5223.}
\nref\nobogi{M. J. Duff, G. W. Gibbons, and P. K. Townsend,
``Macroscopic Superstrings As Interpolating Solitons,''
Phys. Lett. {\bf B332} (1994) 321.}
\nref\gibto{G. W. Gibbons, G. T. Horowitz, and P. K. Townsend,
``Higher Dimensional Resolution Of Dilatonic Black Hole Singularities,''
Class. Quant. Grav. {\bf 12} (1995) 297.}
\nref\gibtot{S. Ferrara, G. W. Gibbons, and R. Kallosh, ``Black Holes
And Critical Points In Moduli Space,'' Nucl. Phys. {\bf B500} (1997) 75,
hep-th/9702103; A. Chamseddine, S. Ferrara, G. W. Gibbons, and R. Kallosh,
``Enhancement Of Supersymmetry Near $5-D$ Black Hole Horizon,''
Phys. Rev. {\bf D55} (1997) 3647, hep-th/9610155.}
\refs{\gibo-\gibtot} and investigations
\nref\olderkleb{S. S. Gubser, I. R. Klebanov, and A. W. Peet,
``Entropy And Temperature Of Black 3-Branes,'' Phys. Rev. {\bf D54}
(1996) 3915.} 
\nref\kkleb{I. R. Klebanov, ``World Volume Approach To Absorption By
Nondilatonic Branes,'' Nucl. Phys. {\bf B496} (1997) 231.}
\nref\kik{S. S. Gubser, I. R. Klebanov, A. A. Tseytlin,
``String Theory And Classical Absorption By Three-branes,''
Nucl. Phys. {\bf B499} (1997) 217.}
\nref\guk{S. S. Gubser and I. R. Klebanov, ``Absorption By Branes And
Schwinger Terms In The World Volume Theory,'' Phys. Lett. {\bf B413}
(1997) 41.} 
\nref\str{J. Maldacena and A. Strominger,  ``Semiclassical Decay Of
Near Extremal Fivebranes,''    hep-th/9710014.} 
\refs{\olderkleb - \str}
of scattering from those metrics,
  Maldacena has made \ref\malda{J. Maldacena, ``The Large $N$
Limit Of Superconformal Field Theories And Supergravity,'' hep-th/971120.}
a remarkable suggestion concerning the 
large $N$ limit not of conventional $SU(N)$ gauge theories but
of some of their conformally invariant cousins.  According to this proposal,
the large $N$ limit of a conformally invariant theory in $d$ dimensions
is governed by supergravity (and string theory) on $d+1$-dimensional
 $AdS$ space
(often called $AdS_{d+1}$) times a compact manifold which in the maximally
supersymmetric cases is a sphere.  There
has also been a discussion of the flow to conformal field theory in some
cases \ref\othermalda{N. Itzhaki, J. M. Maldacena, J. Sonnenschein, and
S. Yankielowicz, ``Supergravity And The Large $N$ Limit Of Theories
With Sixteen Supercharges,'' hep-th/9802042.}
\nref\hyun{S. Hyun, ``$U$-Duality Between Three And Higher Dimensional
Black Holes,'' hep-th/9704005; S. Hyun, Y. Kiem, and H. Shin,
``Infinite Lorentz Boost Along The $M$ Theory Circle And
Nonasymptotically Flat Solutions In Supergravities,'' hep-th/9712021.}
\nref\skend{K. Sfetsos and K. Skenderis, ``Microscopic Derivation Of The
Bekenstein-Hawking Entropy Formula For Nonextremal Black Holes,''
hep-th/9711138; H. J. Boonstra, B. Peeters, and K. Skenderis,
``Branes And Anti-de Sitter Space-times,'' hep-th/9801076.}
\nref\klaus{P. Claus, R. Kallosh, and 
A. van Proeyen, ``$M$ Five-brane And Superconformal $(0,2)$ Tensor Multiplet
In Six-Dimensions,'' hep-th/9711161; P. Claus, R. Kallosh, J. Kumar,
P. Townsend, and A. van Proeyen, ``Conformal Theory Of $M2$, $D3$, $M5$,
and $D1$-Branes $+$ $D5$-Branes,'' hep-th/9801206.}
\nref\kall{R. Kallosh, J. Kumar, and A. Rajaraman, ``Special Conformal
Symmetry Of W0rldvolume Actions,'' hep-th/9712073.}
\nref\frons{S. Ferrara and C. Fronsdal, ``Conformal Maxwell Theory As
A Singleton Field Theory On $AdS(5)$, IIB Three-Branes And Duality,''
hep-th/9712239.}
\nref\ohyun{S. Hyun, ``The Background Geometry Of DLCQ Supergravity,''
hep-th/9802026.}
\nref\guna{M. Gunaydin and D. Minic, ``Singletons, Doubletons And $M$
Theory,'' hep-th/9802047.}
and many other
 relevant discussions of branes, field theories, and $AdS$ spaces
\refs{\hyun - \guna}.

An important example to which this discussion applies 
is $\N=4$ super Yang-Mills
theory in four dimensions, with gauge group $SU(N)$ and coupling constant
$g_{YM}$. This theory conjecturally is equivalent to
Type IIB superstring theory on $AdS_5\times \S^5$, with    string
coupling constant $g_{st}$ proportional to
$g_{YM}^2$, $N$ units of five-form flux
on $\S^5$,  and radius of curvature
$(g_{YM}^2N)^{1/4}$.  In the large $N$ limit with $x=g_{YM}^2N$ fixed
but large, the string theory is weakly coupled and supergravity is
a good approximation to it.  So the hope is that for large $N$ and large
$x$, the $\N=4$ theory in four dimensions is governed by the tree
approximation to supergravity.  In some other important examples discussed
in \malda, there is no dimensionless
parameter analogous to $x$, and supergravity should
apply simply if $N$ is large.

The discussion in \malda\ was motivated by consideration of black holes,
which are also likely to suggest future generalizations.
The black holes in question have near-horizon $AdS$ geometries,
and for our purposes it will suffice to work on the $AdS$ spaces.
$AdS$ space has many unusual properties.  It has a boundary
at spatial infinity
(for example, see \ref\hellis{S. W. Hawking and G. F. R. Ellis,
``The Large Scale Structure Of Space-time,'' Cambridge University Press 
(1973).}, pp. 131-4),
as a result of which  quantization
\ref\avis{S. J. Avis, C. J. Isham, and D. Storey,
``Quantum Field Theory In Anti-de Sitter Space-Time,''
Phys. Rev. {\bf D18} (1978) 3565.}
and analysis of stability \nref\freedman{P. Breitenlohner and
D. Z. Freedman, ``Positive Energy In Anti-de Sitter Backgrounds And
Gauged Extended Supergravity,'' Phys. Lett. {\bf 115B} (1982) 197,
``Stability In Gauged Extended Supergravity,'' Ann. Phys. {\bf 144}
(1982) 197.}
\nref\ggg{G. W. Gibbons, C. M. Hull, and N. P. Warner, ``On The
Stability Of Gauged Supergravity,'' Nucl. Phys. {\bf B218} (1983) 173.}
\nref\mex{L. Mezincescu and P. Townsend,
``Stability At A Local Maximum In Higher Dimensional Anti-de Sitter
Space And Applications To Supergravity,'' Ann. Phys. {\bf 160} (1985) 406.}
\refs{\freedman - \mex}
are not straightforward.  As we describe below,  
the boundary $M_d$ of $AdS_{d+1}$ is in fact
a copy of $d$-dimensional Minkowski space  (with some points
at infinity added); the symmetry group $SO(2,d)$ of $AdS_{d+1}$ acts
on $M_d$ as the conformal group.  The fact that $SO(2,d)$ acts
on $AdS_{d+1}$ as a group of ordinary symmetries and on $M_d$ as a group
of conformal symmetries means that  there are two ways to
get a physical theory with $SO(2,d)$ symmetry:
in a relativistic
field theory (with or without gravity) on $AdS_{d+1}$, or in a conformal
field theory on $M_d$.   Conformal free fields on $M_d$ furnish the
``singleton'' representations of $SO(2,d)$; these are small representations,
originally studied by Dirac (for $d=3$), and interpreted in terms of
free field theory on $M_d$ in
\nref\pik{M. Flato and C. Fronsdal, ``Quantum Field Theory Of Singletons:
The Rac,'' J. Math. Phys. {\bf 22} (1981) 1100.}
\nref\nik{ C. Fronsdal, ``The Dirac Supermultiplet,'' Phys. Rev. {\bf D26}
(1988) 1982.}
\nref\anik{E. Angelopoulos, M. Flato, C. Fronsdal, and D. Sternheimer,
``Massless Particles, Conformal Group, and De Sitter Universe,''
Phys. Rev. {\bf D23} (1981) 1278.}\refs{\pik-\anik}.  
The possible relation of field
theory on $AdS_{d+1}$ to field theory on $M_d$ has been a subject of
long interest; see  
\refs{\frons,\guna} for discussions  motivated by
recent developments, and additional references.

The main idea in \malda\ was not that supergravity, or string theory,
on $AdS_{d+1}$ should be {\it supplemented} by singleton (or other) fields
on the boundary, but that a suitable theory on $AdS_{d+1}$ would
be {\it equivalent} to a conformal field
theory in $d$ dimensions; the conformal field theory might
be described as a generalized singleton theory.   
A precise recipe for computing observables of the conformal field theory
in terms of supergravity on $AdS_{d+1}$ was not given in \malda; obtaining
one
will be the goal of the present paper.  Our proposal
is that correlation functions in conformal field theory are given
by the dependence of the supergravity action on the asymptotic
behavior at infinity. A special case of the proposal is that
dimensions of operators in the conformal field theory are determined
by masses of particles in string theory.  The proposal is effective,
and gives a practical recipe for computing large $N$
conformal field theory correlation functions from supergravity tree diagrams, 
under precisely the conditions proposed in \malda\
-- when the length scale of $AdS_{d+1}$ is large compared to the string and
Planck  scales.

\nref\vann{H. J. Kim, L. J. Romans, and P. van Nieuwenhuizen,
``The Mass Spectrum Of Chiral $N=2$ $D=10$ Supergravity on $\S^5$,
Phys. Rev. {\bf D32} (1985) 389.}
\nref\who{M. Gunaydin and N. Marcus, ``The Spectrum Of The $\S^5$
Compactification Of The Chiral $\N=2$ $D=10$ Supergravity And
The Unitary Supermultiplets Of $U(2,2|4)$,'' Class. Quant. Grav.
{\bf 2} (1985) L11-17.}
One of the most surprising claims in \malda\ was that (for example) to describe
the $\N=4$ super Yang-Mills theory in four dimensions, one should
use not just low energy supergravity on $AdS_5$ but
the whole infinite tower of massive Kaluza-Klein states on $AdS_5\times \S^5$.
We will be able to see explicitly how this works.   Chiral fields in the
four-dimensional $\N=4$ theory (that is, fields in small representations
of the supersymmetry algebra) correspond to Kaluza-Klein harmonics
on $AdS_5\times \S^5$.  Irrelevant, marginal, and relevant perturbations
of the field theory correspond to massive, massless, and ``tachyonic''
modes in supergravity.  The ``tachyonic'' modes have negative mass squared,
but as shown in \freedman\ do not lead to any instability.  The spectrum
of Kaluza-Klein excitations of $AdS_5\times \S^5$, as computed in
\refs{\vann,\who}, can be matched precisely
with certain operators of the $\N=4$
theory, as we will see in section 2.6.  Stringy
excitations of $AdS_5\times \S^5$ correspond to operators whose
dimensions  diverge for $N\to\infty$ in the large $x$ limit.

\nref\otherthooft{G. 't Hooft, ``Dimensional Reduction In Quantum Gravity,''
in {\it Salamfest 1993}, p. 284, gr-qc/9310026.}
\nref\susskind{L. Susskind, ``The World As A Hologram,'' J. Math. Phys.
{\bf 36} (1995) 6377.} 
\nref\bfss{T. Banks, W. Fischler, S. H. Shenker, and
L. Susskind, ``$M$ Theory As A Matrix Model: A Conjecture,'' Phys. Rev.
{\bf D55} (1997) 5112.}
Now we will recall how Minkowski space appears as the boundary
of $AdS$ space.
The conformal group $SO(2,d)$ does not act on Minkowski space, because
conformal transformations can map an ordinary point to infinity.
To get an action of $SO(2,d)$, one must 
add some ``points at infinity.''
A compactification on which $SO(2,d)$ does act is the ``quadric,'' 
described by coordinates $u,v$, $x^1$, $\dots, $ $x^d$, obeying an
equation
\eqn\totally{uv-\sum_{i,j}\eta_{ij}x^ix^j=0,}
and subject to  an overall scaling equivalence ($u\to s u$, $v\to s v$,
$x^i\to sx^i$, with real non-zero 
$s$).  In \totally, $\eta_{ij}$ is the Lorentz
metric of signature $-++\dots +$. \totally\ defines  a  manifold that
admits the action of a group $SO(2,d)$ that preserves the quadratic
form $-du\,dv+\sum_{i,j}dx^idx^j$ (whose signature is $--++\dots +$).
  To show that \totally\ describes 
a compactification of Minkowski space, one notes that generically $v\not= 0$,
in which case one may use the scaling relation to set $v=1$, after
which one uses the equation to solve for $u$.  This leaves
the standard Minkowski space coordinates $x^i$, which parametrize
the portion of the quadric with $v\not= 0$.   The quadric differs
from Minkowsi space by containing also ``points at infinity,'' with
 $v=0$.  The compactification \totally\ is topologically
$(\S^1\times \S^{d-1})/\Z_2$ (the $\Z_2$ acts by a $\pi$ rotation of the first
factor and multiplication by $-1$ on the second),
\foot{After setting $u=a+b$, $v=a-b$, and renaming the variables
in a fairly obvious way, the equation becomes
$a_1^2+a_2^2=\sum_{j=1}^dy_j^2$.  Scaling by only positive $s$,
one can in a unique way map to the locus $a_1^2+a_2^2=\sum_{j=1}^dy_j^2=1$,
which is a copy of $\S^1\times \S^{d-1}$; scaling out also the transformation
with $s=-1$ gives $(\S^1\times \S^{d-1})/\Z_2$.}
 and has closed timelike curves, so one may prefer
to replace it by its universal cover, which is topologically 
$\S^{d-1}\times \R$
(where $\R$ can be viewed as the ``time'' direction).

$AdS_{d+1}$ can be described by the same coordinates $u,v,x^i$ 
but with a  scaling equivalence only under an overall sign change
for all variables, and with \totally\ replaced by
\eqn\botally{uv-\sum_{i,j}\eta_{ij}x^ix^j=1.}
This space is not compact.  If $u,v,x^i$ go to infinity while
preserving \botally, then in the limit, after dividing the coordinates
by a positive constant factor, one gets a solution of \totally.\foot{
Because the constant factor here  is positive, we must identify
the variables in \botally\ under an overall sign change, to get a manifold
whose boundary is $M_d$.}
This is why
the conformal compactification $M_d$ of Minkowski space is the boundary
of $AdS_{d+1}$.  Again, in \botally\ there are closed timelike curves;
if one takes the universal cover to eliminate them, then the boundary
becomes the universal cover of $M_d$.  

According to \malda, an $\N=4$ theory formulated on $M_4$ is equivalent
to Type IIB string theory on $AdS_5\times \S^5$.  We can certainly
identify the $M_4$ in question with the boundary of $AdS_5$; indeed
this is the only possible $SO(2,4)$-invariant relation between these
two spaces. 
The correspondence between $\N=4$ on $M_4$ and Type IIB on 
$AdS_5\times \S^5$ therefore expresses a string theory on $AdS_5\times \S^5$
in terms of a theory on the boundary.
This correspondence is thus ``holographic,'' in the sense of
\refs{\otherthooft, \susskind}. 
This realization of holography is somewhat different from what is obtained
in the matrix model of $M$-theory \bfss, since for instance it is 
covariant (under $SO(2,d)$).  But otherwise the two are strikingly similar.
In both approaches,
$M$-theory or string theory on a certain background is described
in terms of a field theory with maximal supersymmetry. 

The realization of holography via $AdS$ space
is also reminiscent of the relation \ref\witten{E. Witten, ``Quantum Field
Theory And The Jones Polynomial,'' Commun. Math. Phys. {\bf 121} (1989) 
351.} between
conformal field theory in two dimensions and Chern-Simons gauge theory
in three dimensions.  In each case, conformal field theory on a $d$-manifold
$M_d$ is related to a generally covariant theory on a $d+1$-manifold
$B_{d+1}$ whose boundary is $M_d$.  The difference is that in the Chern-Simons
case, general covariance is achieved by considering a field theory
that does not require a metric on spacetime, while 
in $AdS$ supergravity,    general covariance is
achieved in the customary way, by integrating over metrics. 

After this work was substantially completed, I learned of independent
work \ref\kleb{S. S. Gubser, I. R. Klebanov, and A. M. Polyakov,
``Gauge Theory Correlators From Noncritical String Theory,''
hep-th/9802109.} in which a very similar
understanding of the the CFT/$AdS$ correspondence to what we describe
in section 2 is developed,
as well as two papers \nref\ferfr{S. Ferrara and C. Fronsdal,
``Gauge Fields As Composite Boundary Excitations,'' hep-th/9802126.}
\nref\horoog{G. T. Horowitz and H. Ooguri, ``Spectrum Of Large $N$
Gauge Theory From Supergravity,'' hep-th/9802116.} \refs{\ferfr,\horoog}
that consider facts relevant to or aspects of the Hamiltonian
formalism that we consider in section 3.

\newsec{Boundary Behavior }

\subsec{Euclidean Version Of $AdS_{d+1}$}

So far we have assumed Lorentz signature, but
the identification of the boundary of $AdS_{d+1}$ with $d$-dimensional
Minkowski space holds also with Euclidean signature, and it will
be convenient to formulate the present
paper in a Euclidean language.  The Euclidean version of $AdS_{d+1}$
can be described in several equivalent ways.  Consider a Euclidean
space $\R^{d+1}$ with coordinates $y_0,\dots, y_d$, and
let $B_{d+1}$ be the open unit ball, $\sum_{i=0}^dy_i^2<1$.
$AdS_{d+1}$ can be identified as $B_{d+1}$ with the metric
\eqn\rufu{ds^2={4\sum_{i=0}^d dy_i^2\over (1-|y|^2)^2}.}
We can compactify $B_{d+1}$ to get the closed unit ball $\bar B_{d+1}$,
defined by $\sum_{i=0}^d y_i^2\leq 1$.  Its boundary is the sphere
$\S^d$, defined by $\sum_{i=0}^d y_i^2=1$.  
$\S^d$ is the Euclidean version of the conformal compactification of
Minkowski
space, and the fact that $\S^d$ is the boundary
of $\bar B_{d+1}$ is the Euclidean version of the statement that  Minkowski
space is the boundary of $AdS_{d+1}$.
The metric \rufu\ on $B_{d+1}$
does not extend over $\bar B_{d+1}$, or define a metric on $\S^d$, because
it is singular at $|y|=1$.
To get a metric which does extend over $\bar B_{d+1}$, one can pick
a function $f$ on $\bar B_{d+1}$ which is positive on $B_{d+1}$ and has
a first order zero on the boundary (for instance, one can take $f=1-|y|^2$),
and replace $ds^2$ by
\eqn\hgu{d\tilde s^2=f^2 ds^2.}
Then $d\tilde s^2$ restricts to a metric on $\S^d$.  As there is no natural
choice of $f$, this metric is only well-defined up to conformal transformations.
One could, in other words, replace $f$ by 
\eqn\hgo{f\to f e^w}
with $w$ any real function on $\bar B_{d+1}$, and this would induce 
the conformal transformation 
\eqn\obbo{d\tilde s^2\to e^{2w}d\tilde s^2}
in the metric  of $\S^d$.
Thus, while $AdS_{d+1}$ has (in its Euclidean version) a metric
invariant under $SO(1,d+1)$, the boundary $\S^d$ has only a conformal
structure, which is preserved by the action of $SO(1,d+1)$.
\foot{This construction is the basic idea of Penrose's method
of compactifying spacetime by introducing conformal infinity;
see \ref\pr{R. Penrose and W. Rindler, {\it Spinors and Spacetime}, vol. 2
(Cambridge University Press, 1986), chapter 9.}.}

Alternatively, with the substitution $r=\tanh (y/2)$, one can  put the
$AdS_{d+1}$ metric  \rufu\ in the form
\eqn\hombo{ds^2=dy^2+\sinh^2 y \,\,d\Omega^2}                            
where $d\Omega^2$ is the metric on the unit sphere, and $0\leq y<\infty$.
In this representation, the boundary is at $y=\infty$.  Finally,
one can regard $AdS_{d+1}$ as the upper half space $x^0>0$ in 
a space with coordinates $x^0,x^1,\dots,x^d$, and metric
\eqn\obbo{ds^2={1\over x_0^2}\left(\sum_{i=0}^d (dx^i)^2\right).}
In this representation, the boundary consists of a copy of $\R^d$,
at $x^0=0$, together with a single point $P$ at $x^0=\infty$.
($x^0=\infty$ consists of a single point since the metric in the $x^i$
direction vanishes as $x^0\to\infty$).  Thus, from this point of
view, the boundary of $AdS_{d+1}$ is a conformal compactification of
$\R^d$ obtained by adding in a point $P$ at infinity; this of course
gives a sphere $\S^d$.

\subsec{Massless Field Equations}

Now we come to the basic fact which will be exploited in the present
paper.  We start with the case of massless field equations, where the
most elegant statement is possible.  We will to begin with discuss
simple and elementary equations, but the idea is that similar properties
should hold for the supergravity equations relevant to the proposal in \malda\
and in fact for a very large class of supergravity theories, connected
to branes or not.

\nref\deser{S. Deser and R. I. Nepomechie, ``Gauge Invariance Versus
Masslessness in de Sitter Space,'' Ann. Phys. {\bf 154} (1984) 396.}
\nref\vanni{L. Mezincescu, P. Townsend, and P. van Nieuwenhuizen,
``Stability At A Local Maximum In Higher Dimensions And The Definition
Of Masslessness In  AdS In Seven Dimensions,'' Phys. Lett. {\bf 143B} (1984)
384.}
The first case to consider is a scalar field $\phi$.   For such
a field, by the massless field equation we will mean the most
naive Laplace equation $D_iD^i\phi=0$.  (From some points of view, as 
for instance in \refs{\freedman, \deser,\vanni},
``masslessness'' in $AdS$ space requires adding a constant to the equation,
but for our purposes we use the naive massless equation.)  
A basic fact about $AdS_{d+1}$ is that
given any function $\phi(\Omega)$ on the boundary $\S^d$, there is
a unique extension of $\phi$ to a function on $\bar B_{d+1}$ that
has the given boundary values and obeys the field equation.
Uniqueness depends on the fact that there is no nonzero square-integrable
solution of the Laplace equation (which, if it existed, could be added to
any given solution, spoiling uniqueness). This is true because
 given a square-integrable
solution, one would have
by integration by parts
\eqn\roggo{0=-\int_{B_{d+1}}d^{d+1}y \,\,\sqrt g\phi D_iD^i\phi
=\int_{B_{d+1}}d^5y\sqrt g |d\phi|^2,}
so that $d\phi=0$ and hence (for square-integrability) $\phi=0$.  
Now, in the representation \hombo\ of $ B_{d+1}$, the Laplace equation reads
\eqn\nuggo{\left(-{1\over 
(\sinh y)^d}{d\over dy}(\sinh y)^d{d\over dy} +{L^2\over
\sinh^2 y}\right)\phi = 0,}
where $L^2$ (the square of the angular momentum) is the angular
part of the Laplacian.
If one writes $\phi=\sum_\alpha \phi_\alpha(y)f_\alpha(\Omega)$,
where the $f_\alpha$ are spherical harmonics, then the equation for
any $\phi_\alpha$ looks for large $y$ like 
\eqn\bruggo{{d\over dy}e^{dy}{d\over dy}\phi_\alpha=0,}
with the two solutions $\phi_\alpha\sim 1$ and $\phi_\alpha\sim e^{-dy}$.
One linear combination of the two solutions is smooth near $y=0$;
this solution has a non-zero constant term at infinity (or one would get a
square-integrable solution of the Laplace equation).  So for every partial
wave, one gets a unique solution of the Laplace equation with a given constant
value at infinity; adding these up with suitable coefficients, one
gets a unique solution of the Laplace equation with any desired limiting
value $\phi(\Omega)$ at infinity.  In section 2.4, we give
an alternative proof of existence of $\phi$ based on an integral formula.
This requires the explicit form of the metric on $B_{d+1}$ while the
proof we have given is valid for any spherically symmetric metric
 with
a double pole on the boundary.  (With a little more effort, one can make
the proof without assuming spherical symmetry.)

In the case of gravity, by the massless field equations, we mean the Einstein
equations (with a negative cosmological constant as we are on $AdS_{d+1}$).
Any metric on $B_{d+1}$ that has the sort of boundary behavior seen in
\rufu\ (a double pole on the boundary) induces as in \hgu\ a conformal
structure on the boundary $\S^d$.  Conversely, by a theorem of Graham and
Lee \ref\graham{R. Graham and J. Lee, ``Einstein Metrics With Prescribed
Infinity On The Ball,'' Adv. Math. {\bf 87} (1991) 186.}
(see 
\nref\lebrun{C. LeBrun, ``$H$-Space With A Cosmological Constant,''
Proc. Roy. London Ser. A {\bf 380} (1982) 171.}
\nref\fefferman{C. Fefferman and C. R. Graham, ``Conformal Invariants,''
{\it Elie Cartan et les Math\'ematiques d'aujourdhui}''
(Asterisque, 1985), 95.}
\nref\ped{H. Pedersen, ``Einstein Metrics, Spinning Top Motion,
And Monopoles,'' Math. Ann. {\bf 274} (1986) 35.}
\nref\cy{S. Y. Cheng and S.-T. Yau, ``On The Existence Of A Complete
Kahler Metric On Noncompact Complex Manifolds And The Regularity Of
Fefferman's Equation,'' Comm. Pure Appl. Math {\bf 33} (1980) 507.}
\refs{\lebrun - \cy}
for earlier mathematical
work), any conformal structure on $\S^d$
that is sufficiently close to the standard one arises, by the procedure
in \hgu, from a unique metric on $B_{d+1}$ that obeys the Einstein
equations with negative cosmological constant and 
has a double pole at the boundary.  (Uniqueness of course means
uniqueness up to diffeomorphism.)
This is proved by first showing existence and uniqueness at the linearized
level (for a first order deformation from the standard conformal structure
on $\S^d$) roughly along the lines of the above treatment of scalars, 
and then going to small but finite perturbations via an implicit
function argument.

For a Yang-Mills field $A$ with curvature $F$, 
by the massless field equations we mean the usual
minimal Yang-Mills equations $D^iF_{ij}=0$, 
or  suitable generalizations, for instance resulting from adding
a Chern-Simons term in the action.  One expects an analog
of the Graham-Lee theorem stating that any $A$ on $\S^d$ that is sufficiently
close to $A=0$ is the boundary value of a solution of the Yang-Mills
equations on $B_{d+1}$ that is unique up to gauge transformation.  
To first order around $A=0$, one needs  to solve Maxwell's equations
with given boundary values; this can be analyzed as we did for scalars, 
or by an explicit
formula that we will write in the next subsection. The argument
beyond that is hopefully similar to what was done by Graham and Lee.
 For topological
reasons, existence and uniqueness of a gauge field on $B_{d+1}$ with given
boundary values can only hold for $A$ sufficiently close to zero.
\foot{This is proved as follows.  We state the proof for
 even $d$.  (For odd $d$, one replaces $\S^2$ by $\S^1$ and
$c_{\half d+1}$ by $ c_{\half(d+1)}$ in the following argument.)
Consider on $\S^2\times \S^d$ a gauge field with a nonzero $\half d+1^{th}$
Chern class, so that  $\int_{\S^2\times\S^d}\Tr F\wedge F\wedge 
\dots \wedge F\not=0$.  If every gauge field on $\S^d$ could be extended
over $B_{d+1}$ to a solution of the Yang-Mills equations unique up to gauge
transformation, then by making this unique extension fiberwise, one
would get an extension of the given gauge field on $\S^2\times \S^d$
over $\S^2\times B_{d+1}$.  Then one would deduce that
$0=\int_{\S^2\times B_{d+1}}d\Tr F\wedge F\wedge \dots \wedge F
=\int_{\S^2\times \S^d}\Tr F\wedge F\wedge \dots \wedge F$, which is
a contradiction.}  
Likewise in the theorem of Graham and Lee, the restriction to
conformal structures that are sufficiently close to the standard one
is probably also necessary for most values of $d$ (one would prove this
as in the footnote using a family of $\S^d$'s that cannot be extended to
a family of $\bar B_{d+1}$'s).

\subsec{Ansatz For The Effective Action}

We will now attempt to make more precise the conjecture \malda\
relating field theory on the boundary of $AdS$ space
to supergravity (and string theory) in bulk.
We will make an ansatz whose justification, initially, is that
it combines the ingredients at hand in the most natural way.
Gradually, further evidence for the ansatz will emerge.  

Suppose that, in any one of the examples in \malda, one has a massless
scalar field $\phi$ on $AdS_{d+1}$,\foot{In each of the cases considered
in \malda, the spacetime is really $AdS_{d+1}\times W$ for some
compact manifold $W$. At many points in this paper,
 we will be somewhat informal
and suppress $W$ from the notation.} obeying
the simple Laplace equation $D_iD^i\phi=0$, with no mass term or
curvature coupling, or some other equation with the same basic
property: the existence of a unique solution on $\bar B_{d+1}$ with
any given boundary values.

Let $\phi_0$ be the restriction of $\phi$ to the boundary of $AdS_{d+1}$.
We will assume that in the correspondence between $AdS_{d+1}$ and conformal
field theory on the boundary, $\phi_0$  should
be considered to couple to a conformal
field ${\cal O}$, via a coupling $\int_{\S^d}\phi_0{\cal O}$.
This assumption is natural given the relation of the conjectured
CFT/$AdS$ correspondence 
to analyses 
\refs{\olderkleb-\guk}
of interactions of fields with
branes.
$\phi_0$ is conformally invariant -- it has conformal weight zero --
since the use of a function $f$ as in \hgu\ to define 
 a metric on $\S^d$
does not enter at all in the definition of $\phi_0$.  
So conformal invariance dictates that  ${\cal O}$ must
have conformal dimension $d$.  We would like to compute the correlation
functions $\langle {\cal O}(x_1){\cal O}(x_2)\dots {\cal O}(x_n)\rangle$
at least for distinct points $x_1,x_2,\dots, x_n\in \S^n$.  Even better,
if the singularities in the operator product expansion on the boundary
are mild enough, we would like to extend the definition of the correlation
function (as a distribution with some singularities) to allow the points
to coincide.
Even more optimistically, one would like to define the generating functional 
$\langle \exp(\int_{\S^d}\phi_0{\cal O})\rangle_{CFT}$ (the expectation
value of the given exponential in the conformal field theory on the boundary),
for a function $\phi_0$,
at least in an open neighborhood of $\phi_0=0$,
 and not just as a formal series in powers of $\phi_0$.
Usually, in quantum field theory one would face very difficult problems
of renormalization in defining such objects.  We will see
 that
in the present context, as long as one only considers massless
fields in bulk, one meets operators for which the short distance
singularities are so restricted that the expectation value of the exponential
can be defined nicely.

Now, let $Z_S(\phi_0)$
be the supergravity (or string) partition function on $B_{d+1}$
computed with the boundary condition that  at infinity $\phi$ approaches a given
function $\phi_0$.  For example, in the approximation
of classical supergravity, one computes $Z_S(\phi_0)$ by simply extending 
$\phi_0$ over $B_{d+1}$ as a solution $\phi$
of the classical supergravity  equations, and then writing
\eqn\juggo{Z_S(\phi_0) = \exp(-I_S(\phi)),}
where $I_S$ is the classical supergravity
action.  If classical supergravity is not an adequate approximation,
then one must include string theory corrections to $I_S$ (and the equations
for $\phi$), or include quantum
loops (computed in an expansion around the solution $\phi$) rather than
just evaluating the classical action.  Criteria under which stringy
and quantum corrections are small are given in \malda.  
The formula \juggo\ makes sense unless there are infrared divergences
in integrating over $AdS_{d+1}$ to define the classical action $I(\phi)$.
We will argue in section 2.4
 that when such divergences arise, they correspond to
expected renormalizations and anomalies of the conformal field theory.

Our ansatz for the precise relation of conformal field theory on the
boundary to $AdS$ space is that
\eqn\ubbb{\left\langle\exp\int_{\S^d}\phi_0{\cal O}\right\rangle_{CFT}
=Z_S(\phi_0).}
As a preliminary check, note that in the case of 
of $\N=4$ supergravity in four dimensions,
this has the expected scaling of
the 't Hooft large $N$ limit, since the supergravity action $I_S$ that
appears in \juggo\ is of order $N^2$.\foot{The action contains an Einstein
term, of order $1/g_s^2$.  This is of order $N^2$ as $g_{st}\sim g_{YM}^2\sim
1/N$.  It also contains a term $F^2$ with $F$ the Ramond-Ramond five-form
field strength; this has no power of $g_{st}$ but is of order $N^2$
as, with $N$ units of RR flux, one has $F\sim N$.}

Gravity and gauge theory can be treated similarly.  Suppose that
one would like to compute correlation functions of a product of stress
tensors in the conformal field theory on $\S^d$.  The generating
functional of these correlation functions would be the partition
function of the conformal field theory as a function of the metric
 on $\S^d$. 
Except for a $c$-number anomaly, which we discuss in the next subsection
but temporarily suppress, this partition function only depends on the
conformal structure of $\S^d$.   Let us denote as
$Z_{CFT}(h)$ the partition function of the conformal field theory
formulated on a four-sphere with conformal structure $h$.
We interpret the CFT/$AdS$ correspondence to be that 
\eqn\hibbo{Z_{CFT}(h)=Z_S(h)}
where $Z_S(h)$ is the supergravity (or string theory) partition function
computed by integrating over metrics that have a double pole near the boundary
and induce, on the boundary, the given conformal structure $h$.
In the approximation of classical supergravity, one computes $Z_S(h)$
by finding  a solution $g$ of the
Einstein equations with the required boundary behavior, and setting
$Z_S(h)=\exp(-I_S(g))$.  

Likewise, for gauge theory, 
suppose the $AdS$ theory has a gauge     group $G$, of dimension $s$,
with gauge fields $A^a$, $a=1,\dots, s$.
Then in the scenario of \malda, the group $G$ is a global symmetry group
of the conformal field theory on the boundary, and there are currents
$J_a$ in the boundary theory.  We would like to determine the correlation
functions of the $J_a$'s, or more optimistically, the expectation value of the
generating function $\exp(\int_{\S^d} J_aA_0^a)$, with $A_0$ an arbitrary source.
For this we make an ansatz precisely along the above lines:
we let $Z_S(A_0)$ be the supergravity or string theory partition function 
with the boundary condition that the gauge field $A$ approaches $A_0$
at infinity; and we propose that
\eqn\ubbu{\left\langle \exp(\int_{\S^d} J_a A_0^a)\right\rangle_{CFT}
=Z_S(A_0).}
In the approximation of classical field theory, $Z_S(A_0)$ is computed
by extending $A$ over $B_{d+1}$ as a solution of the appropriate equations
and setting $Z_S(A_0)=\exp(-I_S(A))$.

These examples hopefully make the general idea clear.  One computes
the supergravity (or string) partition function as a function of the
boundary values of the massless fields, and interprets this as a generating
functional of conformal field theory correlation functions, for operators
whose sources are the given boundary values.  Of course, in general,
one wishes to compute the conformal field theory partition function
with all massless fields turned on at once -- rather than considering
them separately, as we have done for illustrative purposes.  One also
wishes to include massive fields, but this we postpone until after
performing some illustrative computations in the next subsection.

\bigskip\noindent{\it A Small Digression}

At this point, one might ask what is the significance of the fact
that the Graham-Lee theorem presumably fails for conformal structures
that are sufficiently far from the round one, and that the analogous
theorem for gauge theory definitely fails for sufficiently strong gauge
fields on the boundary.  For this discussion, we will be more specific
and consider what is perhaps the best understood example considered in
\malda, namely the $\N=4$ theory in four dimensions.  One basic
question one should ask is why, or whether, the partition function
of this conformal field theory on $\S^4$ should converge.  On $\R^4$
this theory
 has a moduli space of classical vacua, parametrized by expectation
values of six scalars $X^a$ in the adjoint representation (one requires
$[X^a,X^b]=0$ for vanishing energy).  Correlation functions on $\R^4$ are
not unique, but depend on a choice of vacuum.  On $\S^4$, because of the 
finite volume, the vacuum degeneracy and non-uniqueness
do not arise.  Instead, one should worry that the path integral would
not converge, but would diverge because of the   integration over the flat
directions in the potential, that is over the noncompact
space of constant (and commuting)
$X^a$.  What saves the day is that the conformal
coupling of the $X^a$ to $\S^4$ involves, for conformal invariance,
a curvature coupling $(R/6)\Tr X^2$ ($R$ being the scalar curvature on $\S^4$).  
If one is sufficiently close to the
round four-sphere,  then $R>0$, and the curvature coupling
 prevents the integral
from diverging in the region of large constant $X$'s.  If one is
very far from the standard conformal structure, so that $R$ is negative,
or if the background gauge field
$A_0$ is big enough, the partition function may well diverge.

Alternatively, it is possible that the Graham-Lee theorem, or its
gauge theory counterpart, could fail in a region in which the conformal
field theory on $\S^4$ is still well-behaved.
For every finite $N$, the conformal field theory partition function
on $\S^4$, as long as it  converges sufficiently well, is analytic
as a function of the conformal structure and other background fields.
But perhaps such analyticity can break down in the large $N$ limit;
the failure of existence and uniqueness of the extension of the boundary
fields to such a classical solution could correspond to such nonanalyticity.
Such singularities that arise only
 in the large $N$ limit  
are known to occur in some toy examples 
 \ref\gw{D. J. Gross and E. Witten, ``Possible Third Order Phase Transition In
The Large $N$ Lattice Gauge Theory,'' Phys. Rev. {\bf D21} (1980) 446.},
and in section 3.2 we will discuss an analogous but somewhat different
source of such nonanalyticity.


\subsec{Some Sample Calculations}

We will now carry out some sample calculations, in the approximation
of classical supergravity, to illustrate the above ideas.

First we 
consider an $AdS$ theory that contains a massless scalar $\phi$
with action
\eqn\uggub{I(\phi)= \half\int_{B_{d+1}}d^{d+1}y\sqrt g |d\phi|^2.}
We assume that the boundary value $\phi_0$ of $\phi$  is the source for
a field ${\cal O}$ and that to compute the two point function of ${\cal O}$, 
we must evaluate $I(\phi)$ for a classical solution with boundary
value $\phi_0$.  For this, we must solve for $\phi$ in
terms of $\phi_0$, and then  evaluate the classical action
\uggub\ for the field $\phi$.

\def\x{{\bf x}}
To solve for $\phi$ in terms of $\phi_0$, we first look for a ``Green's 
function,'' a solution $K$ of the Laplace equation on $B_{d+1}$ whose boundary
value is a delta function at a point $P$ on the boundary.  To find this
function, it is convenient to use the representation of $B_{d+1}$ as
the upper half space with metric
\eqn\olpo{ds^2={1\over x_0^2}\sum_{i=0}^d(dx_i)^2,}
and take $P$ to be the point at $x_0=\infty$.   The boundary
conditions and metric are invariant under translations of the $x_i$, so
$K$ will have this symmetry and is a function only of $x_0$.  The
Laplace equation reads
\eqn\noho{{d\over d x_0}x_0^{-d+1}{d\over dx_0}K(x_0)=0.}
The solution that vanishes at $x_0=0$ is
\eqn\gogo{K(x_0)=cx_0^d}
with $c$ a constant.
Since this grows at infinity, there is some sort of singularity
at the boundary point $P$.  To show that this singularity is a delta
function, it helps to make an $SO(1,d+1)$ transformation that maps
$P$ to a finite point.  The transformation
\eqn\ikko{x_i\to {x_i\over x_0^2+\sum_{j=1}^dx_j^2},\,\,\,i=0,\dots,d}
maps $P$ to the origin, $x_i=0$, $i=0,\dots,d$, and transforms
$K$ to
\eqn\ogog{K(x)=c{x_0^d\over (x_0^2+\sum_{j=1}^dx_j^2)^d}.}
A scaling argument shows that $\int dx_1\dots dx_d K(x)$ is independent
of $x_0$; also, as $x_0\to 0$, $K$ vanishes except at $x_1=\dots = x_d=0$.
Moreover $K$ is positive.  So  for $x_0\to 0$, $K$ 
becomes
a delta function supported at $x_i=0$, with unit coefficient if
$c$ is chosen correctly.  Henceforth, we write $\x$ for the $d$-tuple
$x_1,x_2,\dots x_d$, and $|\x|^2$ for $\sum_{j=1}^dx_j^2$.

Using this Green's function, the solution of the Laplace equation
on the upper half space with boundary values $\phi_0$ is
\eqn\nobbo{\phi(x_0,x_i)=c\int d\x'{x_0^d\over
(x_0^2+|\x-\x'|^2)^d} \phi_0(x_i').}
($d\x'$ is an abbreviation for $dx_1'dx_2'\dots dx_d'$.)
It follows that for $x_0\to 0$,
\eqn\jiggo{{\partial\phi\over\partial x_0}\sim dc x_0^{d-1} \int
d\x' {\phi_0(x')\over |\x-\x'|^{2d}}+O(x_0^{d+1}).}

By integrating by parts, one can express $I(\phi)$ as a surface integral,
in fact 
\eqn\humongo{I(\phi)={\rm lim}_{\epsilon\to 0}\int_{T_\epsilon}
d\x\,\sqrt h \,\phi \,(\vec n\cdot \vec\nabla) \phi,}
where $T_\epsilon$ is the surface $x_0=\epsilon$, $h$ is its induced
metric, and $n$ is a unit
normal vector to $T_\epsilon$.  One has $\sqrt h=x_0^{-d}$, $\vec
n\cdot \vec\nabla \phi
=x_0(\partial \phi/\partial
x_0)$.   Since $\phi\to\phi_0$ for $x_0\to 0$, and
$\partial\phi/\partial x_0$ behaves as in \jiggo,
\humongo\ can be evaluated to give
\eqn\tumo{I(\phi)={cd\over 2} \int d\x\,d\x'
{\phi_0(\x)\phi_0(\x')\over |\x-\x'|^{2d}}.}
So the two point function of the operator ${\cal O}$ is a multiple
of $|\x-\x'|^{-2d}$, as expected for a field ${\cal O}$ of conformal
dimension $d$.
\bigskip\noindent
{\it Gauge Theory}

We will now carry out a precisely analogous computation for free $U(1)$
gauge theory.

The first step is to find a Green's function, that is a solution of
Maxwell's equations on $B_{d+1}$ with a singularity only at a single
point on the boundary.  We use again the description of $B_{d+1}$ as
the upper half space $x_0\geq 0$, and we look for a solution of
Maxwell's equations by a one-form $A$ of the form 
$A=f(x_0)\,dx^i$ (for some fixed $i\geq 1$.  We have $dA=f'(x_0)dx^0\wedge 
dx^i$,
so
\eqn\hugp{*dA={1\over x_0^{d-3}}f'(x_0)(-1)^idx^1dx^2\dots \widehat{dx^i}\dots
dx^d,}
where the notation $\widehat{dx^i}$ means that $dx^i$ is to be omitted
from the $d-1$-fold wedge product.  Maxwell's equations $d(*dA)=0$ give
$f(x_0)=x_0^{d-2}$ (up to a constant multiple) and hence we can take
\eqn\jci{A={d-1\over d-2}x_0^{d-2}dx^i,}
where the constant is for convenience.
After the inversion $x_i\to x_i/(x_0^2+|\x|^2)$, we have
$A=((d-1)/(d-2))\left({x_0\over x_0^2+|\x|^2}\right)^{d-2}d\left({x^i\over
x_0^2+|\x|^2}\right).$  We make a gauge transformation, adding to $A$
the exact form obtained as the exterior derivative of
$-(d-2)^{-1}(x_0^{d-2}x_i/(x_0^2+|\x|^2)^{d-1})$,
and get
\eqn\refop{A={x_0^{d-2}dx_i\over (x_0^2+|\x|^2)^{d-1}}-{x_0^{d-3}x_idx_0
\over (x_0^2+|\x|^2)^{d-1}}.}

Now suppose that we want a solution of Maxwell's equations that at $x_0=0$
coincides with $A_0=\sum_{i=1}^da_idx^i$.  Using the above Green's
function, we simply write (up to a constant multiple)
\eqn\hovo{A(x_0,\x)=\int d\x'{x_0^{d-2}\over (x_0^2+|\x-\x'|^2)^{d-1}}a_i(\x')
dx^i-x_0^{d-3}dx_0\int d\x'{(x-x')^ia_i(\x')\over(x_0^2+|\x-\x'|^2)^{d-1}}.}
Hence
\eqn\rovo{\eqalign{F=dA=&
(d-1)x_0^{d-3}dx_0\int d\x'{a_i(\x')dx^i\over (x_0^2+|\x-\x'|^2)^{d-1}}\cr &
-2(d-1)x_0^{d-1}dx_0\int d\x' {a_i(\x')dx^i\over (x_0^2+|\x-\x'|^2)^d}\cr &
-2(d-1)x_0^{d-3}dx_0\int d\x'{(x_i-x'_i)dx^i a_k(\x')(x^k-(x')^k)\over
(x_0^2+|\x-\x'|^2)^d}+\dots,\cr}}
where the $\dots$ are terms with no $dx^0$.

Now, by integration by parts, the action is
\eqn\rucc{I(A)={1\over 2}
\int_{B_{d+1}}F\wedge *F ={1\over 2}\lim_{\epsilon\to 0}\int_{T_\epsilon}
A\wedge *F,}
with $T_\epsilon$ the surface $x_0=\epsilon$.  Using the above formulas
for $A$ and $F$, this can be evaluated, and one gets up to a constant
multiple
\eqn\bucc{I=\int d\x\,d\x'\, a_i(\x)a_j(\x')\left({\delta_{ij}\over
|\x-\x'|^{2d-2}}-{2(x-x')_i(x-x')_j\over |\x-\x'|^{2d}}\right).}
This is the expected form for the two-point function of  a conserved current. 

\bigskip\noindent{\it Chern-Simons Term And Anomaly}

Now let us explore some issues that arise in going beyond the free field
approximation.  We consider Type IIB supergravity on $AdS_5\times \S^5$.
On the $AdS_5$ space there are massless $SU(4)$ gauge fields (which
gauge the $SU(4)$ $R$-symmetry group of the boundary conformal field theory).
The $R$-symmetry is carried by chiral fermions on $\S^4$ (positive
chirality ${\bf 4}$'s and negative chirality $\bar{\bf 4}$'s), so
there is an anomaly in the three-point function of the $R$-symmetry
currents.  Since we identify the effective action of the conformal field theory
with the classical  action of supergravity (evaluated for a classical
solution with given boundary values), the classical supergravity action
must not be gauge invariant.  How does this occur?

\nref\uwho{M. Pernici, K. Pilch, and P. van Nieuwenhuizen, ``Gauged
$N=8$ $d=5$ Supergravity,'' Nucl. Phys. {\bf B259} (1985) 460.}
\nref\awho{M. Gunaydin, L. Romans, and N. Warner, ``Compact And Noncompact
Gauged Supergravity Theories In Five Dimensions,'' Nucl. Phys. {\bf B272}
(1986) 598.}
The classical supergravity action that arises
in $\S^5$ compactification of Type IIB has in addition to the standard 
Yang-Mills action, also a Chern-Simons coupling (this theory
has been described in \refs{\uwho,\awho}; the Chern-Simons term
can be found, for example, in 
eqn. (4.15) in \awho.).  The action is
thus 
\eqn\iccuc{I(A)=\int_{B_5}\Tr\left({ F\wedge *F\over 2g^2}+{iN\over 16\pi^2}
(A\wedge dA\wedge
dA+\dots)\right),}
where the term in parentheses is the Chern-Simons term.  

To determine the conformal field theory effective action for a source
$A_0$, one extends $A_0$ to a field $A$ on $B_{d+1}$ that obeys the classical
equations.  Note that $A$ will have to be complex (because of the $i$ 
multiplying the Chern-Simons term), and since classical equations for
strong complex-valued gauge fields will not behave well,
this is another reason, in
addition to arguments given in section 2.3, that $A_0$ must be sufficiently
small.  Because of gauge-invariance, there is no natural choice of $A$;
we simply pick a particular $A$.

Any gauge transformation on $\S^4$ can be extended over $B_5$;
given a gauge  transformation $g$ on $\S^4$, we pick an arbitrary extension
of it over $B_5$ and call it $\widehat g$.  If $A_0$ is changed by a gauge
transformation $g$, then $A$ also changes by a gauge transformation,
which we can take to be $\widehat g$.

Now we want to see the anomaly in the conformal field theory effective
action $W(A_0)$.  In the classical supergravity approximation, the
 anomaly immediately follows from the relation
$W(A_0)=I(A)$.  In fact, $I(A)$ is not gauge-invariant; under a gauge
transformation
it picks up a boundary term (because the Chern-Simons coupling changes
under gauge transformation by a total derivative) which precisely
reproduces the chiral anomaly on the boundary.  

Going beyond the classical
supergravity approximation does not really change the discussion of the 
anomaly, because for {\it any} $A$ whose boundary values are $A_0$,
the gauge-dependence of $I(A)$ is the same; hence averaging over $A$'s
(in computing quantum loops) does not matter.  Likewise, stringy
corrections to $I(A)$ involve integrals over $B_5$ of gauge-invariant
local operators, and do not affect the anomaly.

\bigskip
\centerline{\vbox{\hsize=4in\tenpoint
\centerline{\psfig{figure=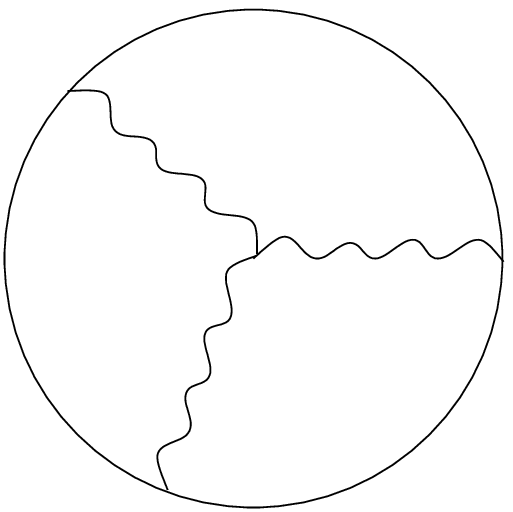}}
\vglue.4in
Fig. 1.  A contribution of order $g$ to the three point function 
$\langle J(P_1)J(P_2)J(P_3)\rangle$ of three currents in the boundary
conformal field theory.   The currents are inserted at points on the
boundary of $AdS_5$ and the interaction takes place in the interior.
The boundary is denoted by a solid circle.  Wavy lines are gluon
propagators, and the vertex in the interior could come from the
conventional Yang-Mills action or from the Chern-Simons term. }}

\bigskip\noindent{\it $n$-Point Function Of Currents}

To actually compute the $n$-point function of currents in the boundary
theory, one would have to compute Feynman diagrams such as those
sketched in Figures 1 and 2.  In the diagrams, the boundary of
$AdS$ space is sketched as a circle; wavy lines are gauge boson propagators.
There are two kinds of propagators.  Propagation between two interior
points is made by the conventional $AdS$  gauge field
propagator.  Propagation between a boundary point and an interior
point is made with the propagator used in \rovo; it expresses the influence
in the interior of a ``source'' on the boundary.
By a ``source'' we mean really a delta function term in the boundary
values.

The operator product coefficients for a product of currents 
would be computed by inserting currents $J_{i_1},\dots J_{i_k}$ at
boundary points $P_1,\dots, P_k$, and letting the $P$'s approach each other,
say at $Q$.  Singularities arise only if some vertices in the interior
of $AdS$ space approach $Q$ at the same time.  After collapsing all 
propagators that connect points that are approaching each other,
one gets a reduced diagram in which some number $q$ of propagators -- with
$q\geq 0$ -- connect the point $Q$ to an interior point.  Intuitively,
one expects this configuration to represent the insertion of an operator
at $Q$.  The OPE coefficient comes from evaluating the collapsing propagators
(and integrating over positions of interaction vertices in the bulk
that are approaching $Q$).
For $q=0$,
the operator that is inserted at $Q$ is a multiple of the identity,
for $q=1$ it is a current $J$, and for general $q$ it appears intuitively
that this operator should correspond to a normal ordered product
of $q$ currents, with derivatives perhaps acting on some of them.
By studying the $AdS$ propagators, it can be shown, at least in simple
cases, that the operator product singularities obtained in this way
are the expected ones.  But we will not try to demonstrate this in the
present paper.

A few noteworthy facts are the following.  {\it Any} $AdS$ theory,
not necessarily connected with a specific supergravity compactification,
appears to give boundary correlation functions that obey the general
axioms of conformal field theory.   If only massless particles are
considered on $AdS$, one apparently
gets an astonishingly simple closed operator product
expansion with only currents (or only currents and stress tensors).
To obtain the more realistic OPE of, for instance, the $\N=4$ super Yang-Mills
theory in four dimensions, one must include additional fields in the
supergravity; we introduce them in sections 2.5 and 2.6.

\centerline{\vbox{\hsize=4in\tenpoint
\centerline{\psfig{figure=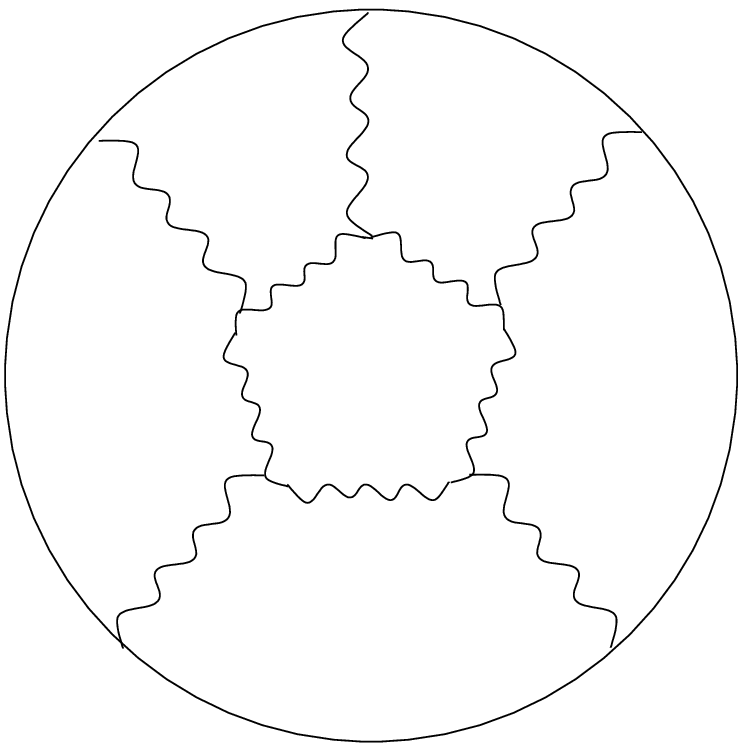}}
\vglue.6in
Fig. 2. A one-loop quantum correction to the five-point function of
currents in the boundary conformal field theory.  }}

\bigskip\noindent{\it Gravity}

Now we will discuss, though only schematically, the gravitational
case, that is, the dependence of the effective action of the boundary
conformal field theory on the conformal structure on the
boundary.   We consider a nonstandard conformal structure on $\S^d$
and find, in accord with the Graham-Lee theorem, an Einstein metric
on $B_{d+1}$ that induces the given conformal structure on $\S^d$.
To compute the partition function of the conformal field theory,
we must evaluate the Einstein action on $B_{d+1}$ for this metric.  In writing
this action, we must remember to include a surface term
\nref\york{J. W. York, ``Role Of Conformal Three Geometry In The Dynamics
Of Gravitation,'' Phys. Rev. Lett. {\bf B28} (1972) 1082.}
\nref\gibhaw{G. W. Gibbons and S. W. Hawking, ``Action Integrals And
Partition Functions In Quantum Gravity,'' Phys. Rev. {\bf B15} (1977) 2752.}
\refs{\york,\gibhaw} in the Einstein action.  
The action thus reads
\eqn\ubbi{I(g)=\int_{B_{d+1}}d^{d+1}x\sqrt g
\left(\half R+\Lambda\right)+\int_{\partial
B_{d+1}}K,}
where $R$ is the 
scalar curvature, $\Lambda$ the cosmological constant, and $K$ is
the trace of the second fundamental form of the boundary.

Now, everything in sight in \ubbi\ is divergent.  The Einstein equations
$R_{ij}-\half g_{ij}R=\Lambda g_{ij}$ imply that $R$ is a constant, so
that the bulk integral in \ubbi\ is just a multiple of the volume of
$B_{d+1}$. This is of course infinite.  Also, there is a question
of exactly what the boundary term in \ubbi\ is supposed to mean, 
if the boundary is at infinity.
To proceed, therefore, we need to regularize the action.  
The rough idea is to pick a positive function $f$ on $B_{d+1}$
that has a first order zero near the boundary.  Of course, this
breaks conformal invariance on the boundary of $B_{d+1}$, 
and determines an actual metric $h$ on $\S^d$ via \hgu.
We can regularize the volume integral by limiting it to the
region $B(\epsilon)$ defined by $f>\epsilon$.  Also, the
boundary term in the action now gets a precise meaning: one integrates
over the boundary of $B(\epsilon)$.  Of course, divergences will appear
as $\epsilon\to 0$.

We want to show at least schematically
that, with a natural choice of $f$, the divergent
terms are local integrals on the boundary.
   The strategy for proving this is as follows.
Theorems 2.1 and 2.3 of \fefferman\ show that the metric
of $B_{d+1}$ is determined locally by the conformal structure of $\S^d$
up to very high order.  The proof of the theorems in section 5 of that paper
involves showing that once one picks a metric on $\S^d$ in its given 
conformal class, one can pick distinguished coordinates near the boundary
of $B_{d+1}$ to very high order.  These distinguished coordinates give
a natural definition of $f$ to high order and with that definition, the
divergent part of the action depends locally on the metric of $\S^d$.
The divergent terms are thus local integrals on $\S^d$,
of the general form $\int_{\S^d}d^dx\sqrt h P(R, \nabla R, \dots)$,
where $P$ is a polynomial in the Riemann tensor of $\S^d$ and its derivatives.

The effective action of the conformal field theory on $\S^d$
can thus be made finite by subtracting local counterterms.
After the subtractions are made, the resulting finite effective action
will not necessarily be conformally invariant, but the conformal
anomaly -- the  variation of the effective action
under conformal transformations of the metric -- will be given by a local
expression.  In fact, the conformal anomaly comes as usual from the
logarithmically divergent term; all the divergent terms are local.

The structure that we have described is just what one obtains when
a conformal field theory is coupled to a background metric.  A regularization
that breaks conformal invariance is required.  After picking a regularization,
one encounters divergent terms which are local; on subtracting them,
one is left with a local conformal anomaly.  The structure of the conformal
anomaly  in general dimensions
is reviewed in \ref\duff{M. J. Duff, ``Twenty Years Of The
Weyl Anomaly,'' Class. Quant. Grav. {\bf 11} (1994) 1387.}.
The conformal anomaly arises only for even $d$; this is 
related to the fact that Theorems 2.1 and 2.3 in \fefferman\ make different
claims for odd and even $d$.  In the even $d$ case, the theorems
allow logarithmic terms that presumably lead to the conformal anomaly.

\subsec{The Massive Case}

We have come about as far as we can considering only massless fields on
$AdS$ space.  To really make contact with the ideas in \malda\ requires
considering massive excitations as well, 
since, among other reasons, the $AdS$ compactifications considered
in \malda\ apparently do not have consistent low energy truncations
in which only massless fields are included.  The reason for this last
statement is that in, for instance, the $AdS_5\times \S^5$ example,
the radius of the $\S^5$ is comparable to the radius of curvature of the
$AdS_5$, so that the inverse radius of curvature (which behaves in $AdS$
space roughly as the smallest wavelength of any excitation, as seen
in \freedman) is comparable to the masses of the Kaluza-Klein excitations.

For orientation, we consider a scalar with mass $m$ in $AdS_{d+1}$.
The action is
\eqn\nogog{I(\phi)= \half\int d\mu \left(|d\phi|^2+m^2\phi^2\right)}
with $d\mu$ the Riemannian measure.
The wave equation, which we wrote before as \nuggo, 
receives an extra contribution from the
mass term, and is now
\eqn\onuggo{\left(-{1\over 
(\sinh y)^d}{d\over dy}(\sinh y)^d{d\over dy} +{L^2\over
\sinh^2 y}+m^2\right)\phi = 0.}
The $L^2$ term is still irrelevant at large $y$.  The two linearly
independent solutions of \onuggo\ behave for large $y$ as $e^{\lambda y}$
where 
\eqn\gonuggo{\lambda(\lambda+d)=m^2.}
For reasons that will be explained later, $m^2$ is limited to the region
in which this quadratic equation
has real roots.  Let $\lambda_+$ and $\lambda_-$ be the larger and smaller
roots, respectively; note that $\lambda_+\geq -d/2$ and $\lambda_-\leq -d/2$.
One linear combination of the two solutions extends smoothly over the interior
of $AdS_{d+1}$; this solution behaves at infinity as $\exp(\lambda_+y)$.

This state of affairs means that we cannot find a solution of the massive
equation of motion \onuggo\ that approaches a constant at infinity.
The closest that we can do is the following.  Pick any positive function
$f$ on $B_{d+1}$ that has a simple zero on the boundary.  For instance,
$f$ could be $e^{-y}$ (which has a simple zero on the boundary, as one
can see by mapping back to the unit ball with $r=\tanh(y/2)$). 
Then one can look for a solution
of the equation of motion that behaves as 
\eqn\juvv{\phi\sim f^{-\lambda_+}\phi_0,}
with an arbitrary ``function'' $\phi_0$ on the boundary.

But just what kind of object is $\phi_0$?  The definition of $\phi_0$
as a function depends on the choice of a particular $f$, which
was the same choice used in \hgu\ to define a metric (and not just a conformal
structure) on the boundary of $AdS_{d+1}$.  If we transform  $f\to e^wf$,
the metric will transform by $d\tilde s^2\to e^{2w}d\tilde s^2$.
At the same time, \juvv\ shows that $\phi_0$ will transform
by $\phi_0\to  e^{w\lambda_+}\phi_0$.  This transformation under
conformal rescalings of the metric shows that $\phi_0$ must be understood
as a conformal density of length dimension $\lambda_+$, or mass dimension
$-\lambda_+$.  (Henceforth, the word ``dimension'' will mean mass dimension.)

\bigskip\noindent
{\it Computation Of Two Point Function}

If, therefore, in a conformal field theory on the boundary of $AdS_{d+1}$, 
there is a coupling $\int\phi_0{\cal O}$ for some operator ${\cal O}$,
then ${\cal O}$ must have conformal dimension $d+\lambda_+$.
Let us verify this by computing the two point function of the field ${\cal O}$.

To begin with, we need to find the explicit form of a function $\phi$
that obeys the massive wave equation and behaves as $f^{-\lambda_+}\phi_0$
at infinity.  We represent $AdS$ space as the half-space $x_0\geq 0$
with metric $ds^2=(1/x_0^2)(dx_0^2+\sum_idx_i^2)$.
As before, the first step is to find a Green's function, that is
a solution of $(-D_iD^i+m^2)K=0$ that vanishes on the boundary except
at one point.  Again taking this point to be at $x_0=\infty$, $K$ should
be a function of $x_0$ only, and the equation reduces to
\eqn\gly{\left(-x_0^{d+1}{d\over dx_0}x_0^{-d+1}
{d\over dx_0}+m^2     \right)K(x_0)=0.}
The solution that vanishes for $x_0=0$ is $K(x_0)=x_0^{d+\lambda_+}$.
After the inversion $x_i\to x_i/(x_0^2+|\x|^2)$, 
we get the solution
\eqn\hurf{K={x_0^{d+\lambda_+}\over (x_0^2+|\x|^2)^{d+\lambda_+}},}
which behaves as $x_0^{d+\lambda_+}$ for $x_0\to 0$ except for
a singularity at $\x=0$.
Now,
\eqn\ikko{\lim_{x_0\to 0}  {x_0^{d+2\lambda_+}
\over(x_0^2+|\x|^2)^{d+\lambda_+}} }
is a multiple of $\delta(\x)$, as one can prove by using a scaling
argument to show that 
\eqn\gikko{\int d\x {x_0^{d+2\lambda_+}
\over(x_0^2+|\x|^2)^{d+\lambda_+}}}
is independent of $x_0$, and observing that for $x_0\to 0$, the function
in \ikko\ is supported near $x=0$.  So if we use the Green's function $K$
to construct the solution
\eqn\jikko{\phi(x)=c'\int d\x'{x_0^{d+\lambda_+}\over
(x_0^2+|\x-\x'|^2)^{d+\lambda_+}}\phi_0(\x')}
of the massive wave equation,
then $\phi$ behaves for $x_0\to 0$ like $x_0^{-\lambda_+}\phi_0(x)$,
as expected.

Moreover, given the solution \jikko, one can evaluate the action
\nogog\ by the same arguments (integration by parts and reduction to
a surface term) that we used in the massless case,
with the result
\eqn\togog{I(\phi)={c'(d+\lambda_+)\over 2}\int d\x\,d\x'
{\phi_0(\x)\phi_0(\x')\over |\x-\x'|^{2(\lambda_++d)}}.}
This is the expected two point function of a conformal field ${\cal O}$
of dimension $\lambda_++d$.

In sum, the dimension $\Delta$ of a conformal field on the boundary
that is related to a field of mass $m$ on $AdS$ space is
$\Delta=d+\lambda_+$, and is the larger root of
\eqn\gbb{\Delta(\Delta-d)=m^2.}
Thus
\eqn\jbb{\Delta=\half(d+\sqrt{d^2+4m^2}).}
In section  3.3, we will propose another explanation, using a Hamiltonian
approach, for this relation.

If one considers not a scalar field but a field with different Lorentz
quantum numbers, then the dimensions are shifted.  For instance, a massless
$p$-form $C$ on $AdS$ space could couple to a $d-p$-form operator
${\cal O}$ on the boundary via a coupling $\int_{M_d}C\wedge {\cal O}$;
conformal invariance (and gauge invariance of the massless $C$ field)
requires that the dimension of ${\cal O}$ be $d-p$.
For $p=1$, ${\cal O}$ is equivalent to a current, and the fact that the
dimension is $d-1$ was verified explicitly in the last subsection.
If instead $C$ is a massive field that behaves as $f^{-\lambda_+}C_0$
near the boundary, then as in our discussion of scalars,
$C_0$ is a $p$-form on the boundary of
conformal dimension $p-\lambda_+$, and an operator ${\cal O}$ that 
couples to $C_0$ has conformal dimension $\Delta=d-p+\lambda_+$.  In this case
\gbb\ is replaced by
\eqn\mbb{(\Delta+p)(\Delta+p-d)=m^2.}

\bigskip\noindent{\it Correlation Functions Of Arbitrary Operators}

Now we would like to go beyond the two-point function and consider
the $n$-point functions of arbitrary operators ${\cal O}_i$ in the
conformal field theory.  Let us first discuss the situation in field
theory.  We can hope to define $n$-point functions $\langle {\cal O}_1(x_1)
{\cal O}_2(x_2)\dots {\cal O}_n(x_n)\rangle$, for distinct points $x_1,\dots,
x_n$.  If one allows arbitrary local operators of any dimension, 
the singularities for coincident points are extremely complicated
and one cannot without a very high degree of complication and arbitrariness
define a  generating function
\eqn\umbo{\left\langle \exp(\int d^dx \,\sum_a\epsilon_a(x) {\cal O}_a(x))
\right\rangle.}
It is nevertheless convenient to formally have the machinery of generating
functionals.  To do this one should consider the $\epsilon_a$ 
to have  disjoint
support and to be infinitesimal, and evaluate \umbo\ only to linear
order in each $\epsilon_a$.  One permits the operators
${\cal O}_a(x)$ to be the same for some distinct values of $a$.

Now let us consider the $AdS$ description.  In doing so, for simplicity
of exposition we suppose that the ${\cal O}_a$ are all scalar operators,
of dimension $d+\lambda_{a,+}$.
They correspond to fields $\phi_a$  on $AdS$ 
which behave near the boundary as $f^{-\lambda_{a,+}}$.  
Now, ideally we would like to claim that \umbo\ equals the supergravity
(or string theory) partition function with boundary conditions such that
$\phi_a$ behaves near the boundary as $f^{-\lambda_{a,+}}\epsilon_a$.
This works fine if the $\lambda_{a,+}$ are all negative; then the
perturbations vanish on the boundary, and the determination of the 
boundary behavior of the fields using the linearized equations is 
self-consistent.  The case of negative $\lambda_{a,+}$ is the case
of relevant perturbations of the conformal field theory, of dimension
$<d$, and it should come as no surprise that this is the
most favorable case for defining the functional \umbo\
 as an honest functional of the $\epsilon_a$ (and not
just for infinitesimal $\epsilon_a$).
One can also extend this to include operators with vanishing
$\lambda_{a,+}$, corresponding to marginal perturbations of the conformal
field theory.  For this, one must go beyond the linearized approximation
in describing the boundary fields, and include general boundary values
of massless fields, which we introduced in section 2.3.

The trouble arises if one wants to include irrelevant (or unrenormalizable)
perturbations of the conformal field theory.  This is the case
in which in field theory one would run into severe trouble in making
sense of the generating functional \umbo, such that one is led 
to consider the $\epsilon_a$ to be infinitesimal and of disjoint support.
One runs into a corresponding phenomenon in the $AdS$ description.
Since the solutions of the linearized equations have
boundary behavior $\phi_a\sim \epsilon_a f^{-\lambda_{a,+}}$,
they diverge near the boundary if $\lambda_+>0$, 
in which case use of the linearized
equations to determine the boundary behavior is not self-consistent.
Just as in the conformal 
field theory, the only cure for this
is to consider the $\epsilon_a$ to be infinitesimal.  
To make sense of the idea of doing the path integral with
boundary conditions $\phi_a\sim \epsilon_af^{-\lambda_{a,+}}$,
we consider the 
$\epsilon_a$ to be not real-valued functions
 but  infinitesimal variables with
$\epsilon_a^2=0$,  and with different $\epsilon_a$ having disjoint support.
We write  $\phi_a= \phi_a'
+\epsilon_af^{-\lambda_{a,+}}$, where $\phi_a'$ is to vanish on the boundary,
and substitute this in the supergravity action (or string theory effective
action).  The substitution is well-behaved, to first order in $\epsilon_a$,
introducing no divergent terms at infinity,
because the shifts are by solutions of the linearized equations. Computing
the path integral over the $\phi_a'$ gives a concrete recipe to compute
the generating functional \umbo\ to the same extent that it makes
sense in field theory.  Of course, in the approximation of classical
supergravity, the path integral is computed just by solving the classical
equations for $\phi_a'$.

\subsec{Comparison To ``Experiment''}

Now we will compare this formalism to ``experiment,'' that is to
some of the surprising features of $AdS$ supergravity.

One very odd feature of the discussion in section 2.5 is that relevant
or superrenormalizable
perturbations correspond to operators with $\lambda_+<0$.  Since
$\lambda_+=(-d+\sqrt{d^2+4m^2})/2$, $\lambda_+$ is negative precisely
if $m^2<0$.
The conformal field theories considered in \malda\ do have relevant 
perturbations.  So the corresponding $AdS$ theories must have
``tachyons,'' that is fields with $m^2<0$.

This seems worrisome, to say the least.  The same worry was faced
in early investigations of gauged and $AdS$ supergravity, where it
was typically found that there were in fact scalar fields of $m^2<0$.
One at first believes that these lead to instabilities, but this proved
not to be the case \refs{\freedman,\mex}.  
Because of the boundary conditions at infinity,
the kinetic energy of a scalar field in $AdS$ space cannot vanish,
and stability requires not that $m^2$ should be positive but that $m^2$
should be no smaller than a certain negative lower bound.

Here is a quick and nonrigorous
Euclidean space derivation of the lower bound.
We will certainly run into a pathology if we consider a scalar field
$\phi$ in $AdS$ space that has a normalizable zero mode, for by
$SO(1,d)$ invariance, there will be infinitely many such modes.
Consider a scalar field whose mass is such that solutions of the
wave equation behave near the boundary (in the representation of $AdS$ by the
metric $dy^2+\sinh^2 y d\Omega^2$)  as $e^{\lambda_+y}$.  Such
a field has no normalizable zero modes if $\lambda_+>-d/2$.  A more
careful analysis shows that there also are no normalizable zero
modes if $\lambda_+=-d/2$.  (This is the case that $\lambda_+=\lambda_-$,
and the solutions of the linearized equation that are regular throughout
$AdS$ space actually behave near infinity as $ye^{-dy/2}$; such solutions
are not normalizable.)  The case $\lambda_+=\lambda_-=-d/2$ arises
for $m^2=-d^2/4$, and this is the lower bound on $m^2$.  If $m^2$ is
smaller, then $\lambda_{\pm }$ become complex, with real part $-d/2$,
and the linearized equations have zero modes with plane wave normalizability;
this is the same sort of spectrum and gives the same sort of pathology
as for tachyon equations in flat Euclidean space.  So the lower
bound is $m^2\geq -d^2/4$, $\lambda_+\geq -d/2$.
Note that the dimension $\Delta=d+\lambda_+$ of a scalar field on the boundary
obtained by the correspondence between boundary
conformal field theory and
$AdS$ thus obeys
\eqn\uvv{\Delta\geq {d\over 2}.}

Making contact with ``experiment'' also requires familiarity with one
other feature of $AdS$ supergravity that caused puzzlement in the old
days.   Consider a theory with maximal supersymmetry -- 32 supercharges --
such as the $AdS_5\times \S^5$ example studied in \malda. 
There are infinitely many massive Kaluza-Klein harmonics.  
Superficially, with 32 supercharges, the spins in any multiplet
should range up to 4.   But supergravity only has fields of spin $\leq 2$!
The resolution of this puzzles involves a subtle $AdS$ analog
\ref\otherfre{D. Z. Freedman and H. Nicolai, ``Multiplet Shortening
In $Osp(N,4)$,'' Nucl. Phys. {\bf B237} (1984) 342} of the construction
\ref\wo{E. Witten and D. Olive, ``Supersymmetry Algebras That Include
Topological Charges,'' Phys. Lett. {\bf 78B} (1978) 97.} 
of small multiplets of flat space
supersymmetry via central charges.  $AdS$ supergravity has special
``small'' multiplets with lower than expected spin, and for
theories with 32 supercharges and only fields of spin $\leq 2$,
{\it all} Kaluza-Klein harmonics are in such reduced multiplets!

The Kaluza-Klein harmonics are important in the present discussion for
the following reason.  In the models considered in \malda, in a limit
in which classical supergravity is valid, the states with masses of order
one are precisely the Kaluza-Klein modes.  Indeed, in for example
the $AdS_5\times \S^5$ model, the radius of the $\S^5$ is comparable
to the radius of curvature of the $AdS_5$ factor, so the Kaluza-Klein
harmonics have masses of order one, in units of the $AdS_5$ length scale.
Stringy excitations are much heavier; for instance, in this model
they have masses of order $(g_{YM}^2N)^{1/4}$ (which is large in the limit
in which classical supergravity is valid), as shown in \malda.

Since the dimension of a scalar operator in the four-dimensional conformal
field theory is
\eqn\mimobo{\Delta= d+\lambda_+={d+\sqrt{d^2+4m^2}\over 2},}
the particles of very large mass in $AdS$ space correspond to operators
of very high dimension in the conformal field theory.  The conformal
fields with dimensions of order one correspond therefore precisely to
the Kaluza-Klein excitations. 

Since the Kaluza-Klein excitations are in ``small'' representations of
supersymmetry, their masses are protected against quantum and stringy
corrections. The conformal fields that correspond to these excitations
are similarly in ``small'' representations, with dimensions
that are protected against quantum corrections.  
It is therefore possible to test the conjectured CFT/$AdS$ correspondence
 by comparing the
Kaluza-Klein harmonics on $AdS_5\times\S^5$ to the  
 operators in the $\N=4$
super Yang-Mills theory that are in small representations 
of supersymmetry; those operators are discussed in 
\ref\seiberg{N. Seiberg, ``Notes On Theories
With Sixteen Supercharges,'' hep-th/9705117.}.

The Kaluza-Klein
harmonics have been completely worked out in \refs{\vann,\who}.
The operators of spin zero are in five infinite families.  We will
make the comparison to the $\N=4$ theory for the three families
that contain relevant or marginal operators.  Two additional families
that contain only states of positive $m^2$ will not be considered here.

We recall the following facts about the $\N=4$ theory. 
This theory has an $R$-symmetry group $SU(4)$, which is a cover of $SO(6)$.
 Viewed as an 
$N=0$ theory, it has six real scalars $X^a$ in the ${\bf 6}$ or vector
of $SO(6)$ and adjoint represention of the gauge group,
 and four fermions $\lambda^A_\alpha$, also
in the adjoint representation of the gauge group, in the ${\bf 4}$ of
$SU(4)$ or positive chirality spinor of $SO(6)$.
  (Here $a=1,\dots, 6$ is a vector
index of $SO(6)$, $A=1,\dots,4$ is a positive chirality $SO(6)$ spinor
index, and $\alpha$ is a positive chirality Lorentz spinor index.)
  From the point of view of an $\N=1$ subalgebra
of the supersymmetry algebra, this theory has three chiral
superfields $\Phi^z$, $z=1,\dots,3$ in the adjoint representation,
and a chiral superfield $W_\alpha$, also in the adjoint representation,
that contains the gauge field strength.  The superpotential is
$W=\epsilon_{z_1z_2z_3}\Tr \Phi^{z_1}[\Phi^{z_2},\Phi^{z_3}]$.
 Viewed as an $\N=2$ theory, this theory
has a vector multiplet and a hypermultiplet in the adjoint representation.

We can now identify the following families of  operators
in the $\N=4$ theory that are in small representations:

(1) First we view the theory from an $\N=1$ point of view.
Any gauge-invariant polynomial in chiral superfields is a chiral
operator.  If it is not a descendant, that is 
it cannot be written in the form $\{\bar Q_{\dot\alpha},
\Lambda^{\dot\alpha}\}$ for any $\Lambda^{\dot\alpha}$,
then it is in a ``small'' representation
of supersymmetry.
Consider the operators $T^{z_1z_2\dots z_k}=
\Tr \Phi^{z_1}\Phi^{z_2}\dots \Phi^{z_k}$.  We require $k\geq 2$
as the gauge group is $SU(N)$ and not $U(N)$.
 If one symmetrizes
in $z_1,\dots,z_k$, one gets chiral fields that 
are in ``small'' representations.  If one does not so symmetrize,
one gets a descendant, because of the fact that the commutators
$[\Phi^{z_i},\Phi^{z_j}]$ are derivatives of the superpotential $W$.
 
 From the $\N=1$ point of view, the theory has a $U(3)$ global
symmetry, of which the $U(1)$ acts as a group of $R$-symmetries.  The
$R$-charge of $T^{z_1z_2\dots z_k}$ is $k$ times that of $\Phi$.
The operator $T^{z_1z_2\dots z_k}$ has dimension $k$ in free field
theory. Its dimension is determined by the $R$-charge and so is exactly
$k$, for all values of the coupling.  From a $U(3)$ point of view,
$T$ transforms in the representation which is the $k$-fold symmetric 
product of the ${\bf 3}$ with itself.  Restoring the $SO(6)$ symmetry
of the $\N=4$ theory, this $U(3)$ representation is part of the $SO(6)$
representation
containing symmetric traceless tensors $T^{a_1a_2\dots a_k}$ of order
$k$.
A field of dimension $k$ in four-dimensional
 conformal field theory has $\lambda_+=k-4$ and corresponds
to a scalar field in supergravity with a mass 
$m^2=\lambda_+(\lambda_++4)=k(k-4)$.
So we expect in the Kaluza-Klein spectrum of the $\N=4$ theory
that for every $k=2,3,\dots$ there should be a scalar with mass
$m^2=k(k-4)$ transforming in the $k^{th}$ traceless symmetric tensor 
representation of $SO(6)$.  These states can be found in eqn. (2.34) and
in Fig. 2 and Table III of \vann.  (Formulas given there contain a parameter
$e$ which we have set  to 1.)  Note that a $k=1$ state does not
appear in the supergravity, which confirms that the supergravity is
dual to an $SU(N)$ gauge theory and not to a $U(N)$ theory.\foot{Another
reason that the $U(1)$ gauge field could not be coded in the $AdS_5$ theory
is that it is free, while in the  $AdS_5$ theory everything couples
to gravity and nothing is free.  To describe $U(N)$ gauge theory on
the boundary, one would have to supplement the $AdS_5$ theory with
an explicit $U(1)$ singleton field on the boundary.}
In fact, the $k=2$ state saturates the bound \uvv, and a $k=1$ state
would violate it.
For
$k=2$ we get the relevant operator $\Tr X^aX^b-(1/6)\delta^{ab}\Tr\, X^2$
(however, the operator $\sum_a\,\Tr (X^a)^2$, which is certainly a relevant
operator for weak coupling, is not chiral and so is related to a stringy
excitation and has a dimension of order $(g^2_{YM}N)^{1/4}$ for strong 
coupling).

(2) Likewise, we can make the chiral superfield $V^{z_1\dots z_t}=
\Tr W_\alpha W^\alpha \Phi^{z_1}\Phi^{z_2}\dots \Phi^{z_t}$ for $t\geq 0$.
Again, modulo descendants one can symmetrize in the ordering of
all $t+2$ factors.  These operators have dimension $t+3$ in free field
theory, and again their dimensions are protected by the $R$-symmetry.
If we set $t=0$, we get the relevant operator $\Tr W_\alpha W^\alpha$,
which is a linear combination of the gluino bilinears $\Tr \lambda_\alpha
^A\lambda^{\alpha B}$,
which transform in the ${\bf 10}$ of $SO(6)$ (one can think of this 
representation as 
consisting of self-dual third rank antisymmetric tensors).  
For higher $t$, these states transform in the representation obtained
by tensoring the ${\bf 10}$ with the $t^{th}$ rank symmetric tensors
and removing traces.  Setting $k=t+1$, we hence expect for $k=1,2,3,\dots$
a supergravity harmonic in the representation just stated with mass
\eqn\hoggo{m^2=(k+2)(k-2).}
Such a harmonic is listed in Fig. 2 and in Table III in \vann.

(3) The conformal field theory also contains a special marginal operator,
the derivative of the Lagrangian density with respect to the Yang-Mills coupling.
This operator is in a small representation, since the Lagrangian cannot
be written as an integral over a superspace with 16 fermionic coordinates.
It transforms in the singlet representation of $SO(6)$, and, being
a marginal operator, is related to a supergravity mode with $m^2=0$, in 
fact, the dilaton.  To exhibit this operator as the first in an infinite
series, view the $\N=4$ theory as an $\N=2$ theory with a vector
multiplet and an adjoint hypermultiplet.  Let $a$ be the
complex scalar in the vector multiplet.  Possible Lagrangians for the
vector multiplets, with $\N=2$ supersymmetry and the smallest
possible number of derivatives, are determined by
a ``prepotential,'' which is a holomorphic, gauge-invariant function of
$a$.  A term $\Tr \,a^r$ in the prepotential leads to an operator
$Q_r= \Tr \,a^{r-2}\,F_{ij}F^{ij}+\dots$ which could be added
to the Lagrangian density while preserving $\N=2$ supersymmetry.  These
operators are all in small representations as the couplings coming
from the prepotential  cannot be obtained by integration over all of 
$\N=2$ superspace.  $Q_r$ has dimension $2+r$ in free field theory,
and this dimension is protected by supersymmetry.  Since $a$ is part of
a vector of $SO(6)$, 
$Q_r$ is part of a set of operators transforming in
the $r-2^{th}$ symmetric tensor representation of $SO(6)$.  (In fact,
$Q_r$ can be viewed as a highest weight vector for this representation.)
Hence, if
we set $k=r-2$, we have for $k\geq 0$ an operator of dimension $k+4$
in the $k^{th}$ symmetric tensor representation, corresponding to
Kaluza-Klein harmonics in that representation with mass
\eqn\njob{m^2= k(k+4).}
These states can be found in Fig. 2 and Table III in \vann.

\newsec{Other Spacetimes And Hamiltonian Formalism}

\subsec{General Formalism}

The CFT-$AdS$ correspondence relates conformal field theory
on $\S^d$ to supergravity on $AdS_{d+1}\times W$, where $W$ is a
compact manifold (a sphere in the maximally supersymmetric cases).
What happens if we replace $\S^d$ by a more general compact $d$-manifold
$N$?  The natural intuitive answer is that one should then replace
$AdS_{d+1}$ by a $d+1$-dimensional Einstein
manifold  $X$  with negative cosmological constant.  The
relation between $X$ and $M$ should be just like the relation
between $\S^d$ and $AdS_{d+1}$.  $X$ should have a compactification consisting
of a manifold with boundary $\bar X$, whose boundary points are $M$ and
whose interior points are $X$, and such that the metric on $X$
has a double pole near the boundary.  Then 
the metric on $X$ determines a conformal structure on $M$, just as we
reviewed in section 2.1 in the case of $AdS_{d+1}$ and $\S^d$.  

More generally, instead of a $d+1$-manifold $X$, one might
use a manifold $Y$ of dimension 10 or 11 (depending on whether one is doing
string theory or $M$-theory) that looks near infinity like $X\times W$ for
some Einstein manifold $X$.  Moreover, $Y$ might contain various
branes or stringy impurities of some kind.  These generalizations
are probably necessary, since $M$ might be, for example, a four-manifold
of non-zero signature, which is not the boundary of any five-manifold.
(However, it may be that for such $M$'s, the conformal field theory
partition function diverges for reasons discussed at the end of section
2.3.)
But since the discussion is non-rigorous anyway, we will keep things
simple and speak in terms of a  $d+1$-dimensional Einstein manifold $X$.

Once $X$ is found, how will we use it to study conformal field theory on
$M$?  As in section 2.3, we propose that the conformal field theory
partition function on $M$ equals the supergravity (or string theory) 
partition function on $X$, with boundary conditions given by the conformal
structure (and other fields) on $M$.  
In the approximation of classical supergravity,
we simply solve the classical equations on $X$ with the given boundary
conditions, and write
\eqn\plop{Z_{CFT}(M)=\exp(-I_S(X)),}
where $Z_{CFT}(M)$ is the conformal field theory partition function on $M$
and $I_S$ is the supergravity action.  

In general, there might be several possible $X$'s.  If so, we have no
natural way to pick one, so as is usual in Euclidean quantum gravity,
we replace \plop\ by a sum.  Before writing
the sum, it is helpful to note that in the large $N$ limit, $I_S(X)$ is
proportional to a positive power of $N$, in fact $I_S=N^\gamma F(X)$ for some
$\gamma>0$.  
For example, $\gamma=2$ for $\N=4$ super Yang-Mills in four dimensions.
So we postulate that in general
\eqn\hiffo{Z_{CFT}(M)=\sum_i \exp(-N^\gamma F(X_i)),}
where the $X_i$ are the Einstein manifolds of boundary $M$.

Another refinement involves spin structures.  $M$ may admit several
spin structures, in which case $Z_{CFT}(M)$ will in general depend on the
spin structure on $M$.  If so, we select a spin structure on $M$ and restrict
the sum in \hiffo\ to run over 
those $X_i$ over which the given spin structure on $M$ can be extended.
\foot{There may be further refinements of a similar nature.  For example,
in the case that the boundary theory is four-dimensional  gauge
theory with a gauge group that is locally $SU(N)$, it may be that
the gauge group is really $SU(N)/\Z_N$, in which case the partition function
on $M$ depends on a choice of a ``discrete magnetic flux''  $w\in H^2(M,\Z)$.
Perhaps one should restrict the sum in \hiffo\ to $X_i$ over which $w$
extends.}

Of course, \hiffo\ should be viewed as
 the classical supergravity approximation to an
an exact formula
\eqn\biffo{Z_{CFT}(M)=\sum_iZ_S(X_i),}
where $Z_S(X_i)$ is the partition function of string theory on $X_i$.

Now, as $M$ is compact, there are no conventional phase transitions
in evaluating path integrals on $M$.  As long as $Z_{CFT}(M)$ is
sufficiently convergent (we discussed obstructions to convergence at the
end of section 2.3), $Z_{CFT}(M)$ is a smooth function of the conformal
structure of $M$ (and other fields on $M$).  However, in \hiffo\ we see
a natural mechanism for a singularity or phase transition that would
arise only in the large $N$ limit.  In the large $N$ limit, the sum
in \hiffo\ will be dominated by that $X_i$ for which $F(X_i)$ is smallest.
If $F(M)=-\ln Z_{CFT}(M)$ is the conformal field theory free energy,
then
\eqn\ico{\lim_{N\to \infty}{F(M)\over N^\gamma}=F(X_i),}
for that value of $i$ for which $F(X_i)$ is least.  At a point at
which $F(X_i)=F(X_j)$ for some $i\not= j$, one may well ``jump'' from one
branch to another.  This will produce a singularity of the large $N$ theory,
somewhat similar to large $N$ singularities found \gw\ in certain toy models.

\subsec{A Concrete Example}

We will next describe a concrete example, namely $M=\S^1\times \S^{d-1}$,
in which one can explicitly describe two possible $X$'s and a transition
between them.  In fact,
the two solutions were described (in the four-dimensional case)
by Hawking and Page
\ref\page{S. Hawking and D. Page, ``Thermodynamics Of Black Holes
In Anti-de Sitter Space,'' Commun. Math. Phys. {\bf 87} (1983) 577.}
who also pointed out the existence of the phase transition (which they
of course interpreted in terms of quantum gravity rather than boundary
conformal field theory!).

The first solution is obtained as follows.
The simplest way to find Einstein manifolds is to take the quotient
of a portion of $AdS_{d+1}$ by a group that acts discretely on it.
$AdS_{d+1}$ can be described as the quadric
\eqn\tomo{uv-\sum_{i=1}^dx_i^2=1.}
(The analogous formula for $AdS_{d+1}$ with Lorentz signature is
given in \botally.)
We restrict ourselves to the region $AdS^+_{d+1}$ of $AdS$ space
with $u,v>0$.
We consider the action on $AdS^+_{d+1}$ of a group $\Z$ generated
by the transformation
\eqn\bomo{u\to \lambda^{-1}u,\,\, v\to \lambda v,\,\,x_i\to x_i,}
with $\lambda$ a fixed real number greater than 1.
This group acts freely on $AdS^+_{d+1}$.  Let $X_1$ be the quotient
$X_1=AdS^+_{d+1}/\Z$.  To describe $X_1$,
we note that a fundamental domain for the action of $\Z$ on $v$ is
$1\leq v\leq \lambda$, with $v=1$ and $v=\lambda$ identified.
$v$ parametrizes a circle, on which a natural angular coordinate
is 
\eqn\huffy{\theta=2\pi \ln v/\ln \lambda.}
For given $v$, one can solve for $u$ by $u=(1+\sum_ix_i^2)/v$.
So $X_1$ is spanned by $x_i$ and $\theta$, and is topologically
$\R^4\times \S^1$.  To see what lies at ``infinity'' in $X_1$, we drop
the 1 in \tomo, to get
\eqn\jubo{uv-\sum_{i=1}^dx_i^2=0,}
 and regard $u,v, $ and the $x_i$ as homogeneous coordinates,
 subject to a scaling relation $u\to su $, $v\to sv$, $x_i\to sx_i$
 with real positive $s$ ($s$ should be positive since we are working
 in the region $u,v>0$).
\jubo\ does not permit the $x_i$ to all vanish (since  $u,v>0$), 
so one can use the scaling relation  to set	
$\sum_{i=1}^dx_i^2=1$.  The $x_i$ modulo the scaling relation thus define
a point in $\S^{d-1}$.  Just as before, $v$ modulo the action   of \bomo\
defines a point in $\S^1$, and one can uniquely solve \jubo\ for $u$.
Hence the boundary of $X_1$ is a copy of $M=\S^1\times \S^{d-1}$.

An important property of this example is that it is invariant under an action
of $SO(2)\times SO(d)$.  Here, $SO(d)$ is the rotation group of the second
factor in $M=\S^1\times \S^{d-1}$; it acts by rotation of the $x_i$ in
\tomo.  $SO(2)$ acts by rotation of the angle $\theta$ defined in \huffy.
The $SO(2)\times SO(d)$ symmetry uniquely determines the conformal
structure of $\S^1\times \S^{d-1}$ up to a constant $\beta=\ln \lambda$
which one can think of as the ratio of the radius of $\S^1$ to the
radius of $\S^{d-1}$.  
$M=\S^1\times \S^{d-1}$ has two spin structures -- spinors may be periodic
or antiperiodic around the $\S^1$.  The partition functions of $M$ with
these spin structures can be interpreted respectively
as
\eqn\ubbu{Z_1(M)=\Tr\,e^{-\beta H}(-1)^F}
or 
\eqn\mubbu{Z_2(M)=\Tr\,e^{-\beta H}.}
Here $H$ is the Hamiltonian that generates time translations if one
quantizes on $\S^{d-1}\times \R$ (with $\R$ as the ``time'' direction);
or it is the dilation generator if the theory is formulated on $\R^d$.
Note that $Z_1(M)$ is not really a supersymmetric index, at least not in
a naive sense, since compactification on $\S^{d-1}$ breaks supersymmetry,
that is, the dilation generator $H$ does not commute with
any supersymmetries.

Both spin structures extend over $X_1$, so $X_1$ contributes to 
both $\Tr\, e^{-\beta H}(-1)^F$ and $\Tr \,e^{-\beta H}$ 
in the conformal field theory.
Its contribution is, however, extremely simple in the approximation of
classical supergravity.  The reason is that $X_1=\S^1\times \R^d$,
and $\beta $ enters only as the radius of $\S^1$.  This ensures that the
supergravity action is linear in $\beta $, 
\eqn\goho{I_S(X_1)=\beta E} 
for some constant $E$,
since $I_S(X_1)$ is computed from a local integral on $X_1$ (plus a boundary
term that is a local integral at infinity), and $\beta $ only enters in
determining how far one must integrate in the $\S^1$ direction.
If one introduces additional background fields on $M=\S^1\times \S^{d-1}$
in a way that preserves
the time-translation invariance of $\S^1\times \S^{d-1}$, then $E$
may depend on the background fields, but is still independent of $\beta $.
$E$ should be interpreted as the ground state energy in quantization of the
conformal field theory on $\S^{d-1}$.

If we set $r^2=\sum_ix_i^2$, $\sum_i (dx_i)^2=(dr)^2+r^2d\Omega^2$,
$t=\ln v+\half\ln(1+r^2)$, then the metric of $X_1$, which is just
the $AdS$ metric  $-du\,dv+\sum_i(dx_i)^2$ with some global identifications,
becomes
\eqn\ugin{ds^2=(1+r^2)(dt)^2+{(dr)^2\over 1+r^2}+r^2d\Omega^2,}
which is the form in which it is written in \page.

\bigskip\noindent{\it The Schwarzschild Solution}

The second solution $X_2$ is the $AdS$ Schwarzschild solution.  Topologically
$X_2=\R^2\times \S^{d-1}$.  The $\S^1$ factor in $M=\S^1\times \S^{d-1}$
is the boundary (in a sense with which we are by now familiar)
of the $\R^2$ factor in $X_2$.  Being simply-connected,
$X_2$ has a unique spin structure, which restricts on $\S^1$
to the antiperiodic spin structure.  (This amounts to the familiar
fact in superstring theory that the antiperiodic or Neveu-Schwarz
spin structure on a circle is the one that extends over a disc.)
Hence $X_2$ contributes to $\Tr\, e^{-\beta H}$, but not to
$\Tr\,e^{-\beta H}(-1)^F$.  Hence $\Tr\,e^{-\beta H}$, but not
$\Tr\,e^{-\beta H}(-1)^F$, 
 may have a large $N$ phase transition
from a ``flop'' between $X_1$ and $X_2$.  That it does, at least for
$d+1=4$,  is equivalent
to facts described by Hawking and Page \page. 

The metric of a Euclidean $AdS_{d+1}$ Schwarzschild black hole of mass $m$,
 for $d=3$, as described in \page, eqn. 2.1,
is
\eqn\bhads{ds^2=V(dt)^2+V^{-1}(dr)^2+r^2d\Omega^2}
with $d\Omega^2$ the line element on a round two-sphere, and
\eqn\hvv{V=1-{2m\over r}+{r^2}.}
(Notation compares to that in \page\ as follows:
we set $b=1$, since the form of the $AdS$ metric that we have used
corresponds for $AdS_4$ to $\Lambda=-3$;
we set $m_p=1$, and write  $m$ instead of $M$ for the black
hole mass.)  Let $r_+$ be the largest
root of the equation $V(r)=0$.  The spacetime is limited to the region
$r\geq r_+$.  The metric is smooth and complete if $t$ is regarded as 
an angular variable with period
\eqn\nabs{\beta={12\pi r_+\over 1+3r_+^2}.}
This has a maximum as a function of $r_+$, so
the black hole spacetime $X_2$ contributes to the thermodynamics only
for sufficiently small   $\beta$, that is sufficiently large temperature.

The action difference between $X_2$ and $X_1$ was computed by Hawking
and Page to be
\eqn\hoddo{I={\pi r_+^2(1-r_+^2)\over 1+3r_+^2}.}
This is negative for sufficiently large $r_+$, that is for sufficiently
small $\beta$ or sufficiently large temperature.  
Thus (for $d=3$ and presumably for all $d$), the boundary
conformal field theories that enter the CFT-$AdS$ correspondence have
phase transitions as a function of temperature in the large $N$ limit.

Further support for the interpretation of the $AdS$ black hole in terms
of conformal field theory at high temperature comes from the fact
that the $AdS$ black hole has positive specific heat \page\ in the region
in which it has lower action.
This is needed for the correspondence with conformal field theory (which
certainly has positive specific heat),
and is a somewhat surprising result, as it does not hold for
Schwarzschild black holes in asymptotically flat space.

On the other hand, it is also shown in \page\ that very low mass $AdS$
black holes have negative specific heat, like black holes in 
asymptotically flat space.  In fact, very small mass black holes
have a size much less than the $AdS$ radius of curvature, and are very
similar to flat space black holes.  Such black holes should decay
by emission of Hawking radiation; the CFT/$AdS$ correspondence
gives apparently a completely unitary, quantum mechanical description
of their decay.

\bigskip\noindent{\it Comparison With Gauge Theory Expectations}

We will now specialize to the $AdS_5\times \S^5$ case, and compare
to the expected large $N$ behavior of four-dimensional
gauge theories.

Actually, we do not know how the $\N=4$ theory should behave for large
$N$ with large and fixed $g^2_{YM}N$.  We will therefore compare to the 
expected large $N$ thermodynamics
\ref\thorn{C. Thorn, ``Infinite $N(C)$ QCD At Finite Temperature: Is
There An Ultimate Temperature?'' Phys. Lett. {\bf 99B} (1981) 458.}
of an $SU(N)$ gauge theory with massive quarks in the fundamental
representation (that is, ordinary QCD generalized to large $N$).
It is surprising that this comparison will work, since
QCD is confining, while the $\N=4$ theory
has no obviously analogous phenomenon.
However, the comparison seems to work, as we will see, at least for
large $g_{YM}^2N$.

The whole idea of the $1/N$ expansion as an approach to QCD \thooft\ is
that for large $N$, the Hilbert space of the theory consists
of color singlet particles whose masses and multiplicities are
independent of $N$.  Suppose we formulate the theory
on some three-manifold $K$  and define $Z(\beta)=\Tr \,e^{-\beta H}$,
and $F=-\ln Z$.  The ground state energy $E$ in this theory
is of order $N^2$ (as vacuum diagrams
receive contributions from $N^2$  species of gluon), while excitation energies
and multiplicities are of order one.   Hence, we expect that
\eqn\uccu{\lim_{N\to\infty}{F(\beta)\over N^2}=\beta {E\over N^2}.}
The idea is that as the excitations have multiplicity of order one,
they contribute to $F$ an amount of order one, and make a vanishing
contribution to the large $N$ limit of $F/N^2$, which hence
should simply measure the ground state energy.
The formula in \uccu\
precisely has the structure of the contribution \goho\ of the
manifold $X_1$ to the free energy.

In large $N$ QCD, such behavior cannot prevail at all temperatures.
At sufficiently high temperatures, one ``sees'' that the 
theory is made of quarks and gluons.  By asymptotic freedom, the gluons
are effectively free at high temperatures.  As the number of species
is of order $N^2$, they contribute to the free
energy an amount of order $N^2$, with a temperature dependence
determined by $3+1$-dimensional conformal invariance.  So for high
enough temperature,
$\lim_{N\to\infty}(F(\beta)/N^2)\sim T^4=\beta^{-4}$.

This behavior is precisely analogous to what we have found for the 
large $N$ behavior of the $\N=4$ theory.  The low temperature phase,
in which the $N^2$ term in the free energy is trivial (as it reflects
only the contribution of the ground state), corresponds to the
phase in which the manifold $X_1$ dominates.  The high temperature
phase, with a non-trivial $N^2$ term in the free energy, corresponds
to the phase in which the manifold $X_2$ dominates.

\subsec{Hamiltonian Interpretation}

Our remaining goal will be to develop a Hamiltonian version of the
CFT-$AdS$ correspondence and to use it to give another explanation of the
relation between masses on $AdS$ space and conformal dimensions on
the boundary theory.

We would like to describe in terms of supergravity the Hilbert space
of the boundary conformal field theory on ${\bf S}^{d-1}$.
The partition function of the conformal field theory on $\S^1\times \S^{d-1}$
can be computed in terms of the states obtained by quantizing on $\S^{d-1}$.
We want the Hilbert space relevant to the low temperature phase
-- the low-lying excitations of the large $N$ vacuum.  So we
must compare the conformal field theory to the manifold $X_1=\S^1\times
\R^d$.  We recall that $X_1$ is just $AdS$ with some global identifications,
with metric found in \ugin:
\eqn\bugin{ds^2=(1+r^2)(dt)^2+{(dr)^2\over 1+r^2}+r^2d\Omega^2.}  
A $t=0$ slice of $X_1$ is a copy of $\R^d$ which can be regarded as
an initial value hypersurface in $AdS_{d+1}$.  The Hilbert space obtained
by quantization on this $t=0$ slice is hence simply the quantum Hilbert
space of the theory on $AdS_{d+1}$.  

So we identify the Hilbert space of the conformal field theory in
quantization on $\S^{d-1}$ with the Hilbert space of the supergravity
(or string theory) on $AdS_{d+1}$.  

Now, in conformal field theory in $d$ dimensions, the Hilbert space on
$\S^{d-1}$ has an important interpretation, which is particularly
well-known in the $d=2$ case. Consider formulating
the theory on $\R^d$ and inserting an operator ${\cal O}$ at a point in $\R^d$,
say the origin.  Think of $\S^{d-1}$ as a sphere around the origin.
The path integral in the interior of the sphere determines a quantum
state $|\psi_{\cal O}\rangle$
on the boundary, that is on $\S^{d-1}$.  Conversely, suppose
we are given a quantum state $|\psi\rangle$
on $\S^{d-1}$.  Given the conformal invariance,
the overall size of the $\S^{d-1}$ does not matter.  So we cut out
a tiny hole around the origin, whose boundary is a sphere of radius $\epsilon$,
and use the state $|\psi\rangle$ to define boundary conditions on the
boundary.  In the limit as $\epsilon\to 0$, this procedure defines
a local operator ${\cal O}_\psi$ as seen by an outside observer.
These two operations are inverses of each other and give a natural
correspondence in conformal field theory
between states on $\S^{d-1}$ and local operators.

So we can assert that, in the CFT-$AdS$ correspondence, the local
operators in the boundary conformal field theory correspond to the
quantum states on $AdS_{d+1}$.

As an application of this, we will recover the relation claimed in
section 2.5 between dimensions of operators in conformal field theory
and particle masses on $AdS$ space.
Choosing a point at the origin in $\R^d$ breaks the Euclidean
signature conformal group $SO(1,d+1)$
to a subgroup that contains $SO(1,1)\times SO(d)$.  $SO(d)$ is the group
of rotations around the given point, and 
$SO(1,1)$ is the group of dilatations.  The $SO(1,1)$ eigenvalue is
the dimension of an operator.   When the $AdS$ metric is written as
in \bugin, an $SO(d)$ group (which rotates the sphere whose line
element is $d\Omega^2$) is manifest, and commutes with the symmetry
of time translations. We call the generator of time translations the energy.
The $SO(1,1)$ generator that measures
operator dimensions in the conformal field theory
is the unique $SO(1,d+1)$ generator that commutes
with $SO(d)$, so it is the energy operator of the $AdS$ theory.

Thus, the energy of a mode in $AdS$ space equals the dimension of the 
corresponding operator in the boundary conformal field theory.
Now, in equation (A6) in \freedman, it is shown that a scalar field of
mass $m^2$ in $AdS$ space has a discrete
energy spectrum (with the definition of energy 
that we have just given), and that the         frequency of the mode of lowest
energy is $\omega=\half(d+\sqrt{d^2-4m^2})$.  Comparing to the formula
\jbb\ for the dimension $\Delta$ of the conformal field ${\cal O}$ 
in boundary conformal field theory that is related to a scalar of mass
$m$ in $AdS$ space, we see that $\Delta=\omega$, as expected.
\foot{In \freedman, the computation is performed only for $d=3$.
Also, the parameter $\alpha$
in \freedman\ is our $-m^2$, and there is a small clash in notation:
the quantity $\lambda_+$  defined on p. 276 of \freedman\ is 
$d+\lambda_+$ in our notation.}  The same equation (A6) also shows that
the energies of excited modes of the massive field are of the form
$\omega+{\rm integer}$, and so agree with the dimensions of derivatives
of ${\cal O}$.

\bigskip
This work was supported in part by NSF Grant PHY-9513835.
I am grateful to N. Seiberg for many very helpful comments,
to G. Tian for pointing out the work of Graham and Lee, and
to D. Freedman and P. van Nieuwenhuyzen for discussions of gauged
supergravity.
\listrefs
\end